\documentclass[12pt]{article}

\usepackage{graphicx}
\usepackage{amssymb}
\usepackage{amsmath}
\usepackage{amsbsy}
\usepackage{setspace}
\usepackage{natbib}
\usepackage{bm}
\usepackage{textcomp}
\usepackage{color}
\usepackage{amsthm}
\usepackage{enumitem}
\usepackage{longtable,threeparttablex,booktabs}
\usepackage{xr-hyper}
\usepackage{hyperref}
\usepackage{mathtools}
\usepackage{lscape}
\usepackage{subfigure}
\usepackage{multirow}
\usepackage{epstopdf}
\usepackage{lineno}
\newtheorem{theorem}{Theorem}
\newtheorem{corollary}{Corollary}
\newtheorem{proposition}{Proposition}
\newtheorem{lemma}{Lemma}
\newtheorem{example}{Example}

\newtheorem{definition}{Definition}

\def\text#1{\mbox{\rm #1}}

\def\underwiggle 1{
\ifmmode\setbox\TempBox=\hbox{$ 1$}\else\setbox\TempBox=\hbox{
1}\fi \setbox\TempBoxA=\hbox to \wd\TempBox{\hss\char'176\hss}
\rlap{\copy\TempBox}\smash{\lower9pt\hbox{\copy\TempBoxA}} }

\newcommand{\E}{\mathrm E}

\newcommand{\var} {\mbox{var}}

\newcommand{\beq}{\begin{equation}}
\newcommand{\eeq}{\end{equation}}
\newcommand{\beas}{\begin{eqnarray*}}
\newcommand{\eeas}{\end{eqnarray*}}
\newcommand{\bea}{\begin{eqnarray}}
\newcommand{\eea}{\end{eqnarray}}
\newcommand{\bei}{\begin{itemize}}
\newcommand{\eei}{\end{itemize}}
\newcommand{\ben}{\begin{enumerate}}
\newcommand{\een}{\end{enumerate}}
\newcommand{\bet}{\begin{theorem}}
\newcommand{\eet}{\end{theorem}}
\newcommand{\bel}{\begin{lemma}}
\newcommand{\eel}{\end{lemma}}
\newcommand{\bep}{\begin{proposition}}
\newcommand{\eep}{\end{proposition}}
\newcommand{\bed}{\begin{definition}}
\newcommand{\eed}{\end{definition}}
\newcommand{\bec}{\begin{corollary}}
\newcommand{\eec}{\end{corollary}}
\newcommand{\bex}{\begin{example}}
\newcommand{\eex}{\end{example}}

\begin{document}
\title{Regression analysis of multiplicative hazards model with time-dependent coefficient for sparse longitudinal covariates
}

\author{Zhuowei Sun$^{1,2}$ and Hongyuan Cao$^{3}$\thanks{Corresponding author,  E-mail address: hongyuancao@gmail.com}}

\footnotetext[1]{~ School of Public Health, Dalian Medical University, Dalian, 116044, Liaoning, China.} 
\footnotetext[2]{~ School of Mathematics, Jilin University, Changchun, 130012, Jilin, China.} 
\footnotetext[3]{~ Department of Statistics, Florida State University, Tallahassee, FL 32306, U.S.A.}
 
\date{}
\maketitle 
\begin{abstract}
We study the multiplicative hazards model with intermittently observed longitudinal covariates and time-varying coefficients. For such models, the existing {\it ad hoc} approach, such as the last value carried forward, is biased. We propose a kernel weighting approach to get an unbiased estimation of the non-parametric coefficient function and establish asymptotic normality for any fixed time point. Furthermore, we construct the simultaneous confidence band to examine the overall magnitude of the variation. Simulation studies support our theoretical predictions and show favorable performance of the proposed method. A data set from Alzheimer's Disease Neuroimaging Initiative study is used to illustrate our methodology. 

\end{abstract}

\noindent \textbf{Keywords:}
Kernel weighting; non-parametric regression; simultaneous confidence band; varying coefficient model.

\section{Introduction}\label{intro.sec}
In clinical trials and epidemiological studies, it is of interest to explore the relationship between longitudinally collected covariates and time-to-event outcomes. The celebrated proportional hazards model postulates a multiplicative relationship between covariates and the hazard function:
\begin{equation}\label{equ:cox}
\lambda\{t \mid Z\}=\lambda_0(t)\mbox{exp}\{\beta_0^{\mathrm{T}}Z\},
\end{equation}
where $\lambda_0(t)$ is an unspecified baseline hazard function, $Z \in {\mathbb R}^p$ is the covariate and $\beta_0 \in {\mathbb R}^p$ is the unknown regression coefficient \citep{cox1972}. In (\ref{equ:cox}), it is assumed that the hazard ratio $\beta_0$ is constant. In practice, this assumption can be violated, as demonstrated in \cite{stensrud2020}. To accommodate hazard ratio that varies with respect to time, \cite{tian2005} proposed a model that replaces $\beta_0$ by $\beta_0(t)$ in (\ref{equ:cox}) and developed estimating equations for statistical inference. In \cite{tian2005}, the covariate $Z$ can be time-dependent, denoted as $Z(t),$ and it is assumed that the entire trajectory of $Z(t)$ is available. 
For longitudinally collected $Z(t),$ only intermittent values are available. 

An \textit {ad hoc} approach to deal with longitudinally collected 
$Z(t)$ is the last value carried forward, where the most recently observed longitudinal covariate is imputed as the current value for each subject. First, this assumes that $Z(t)$ does not change from the time of the last measurement. Second, the uncertainty inherent in the imputation is not considered, as the imputed value is treated indiscriminately with observed data. As a result, substantial bias can arise, which leads to erroneous inferences, as shown in \cite{andersen2003, molnar2009, cao2015a, cao2021}.


As an alternative, a joint modeling strategy is commonly adopted \citep{wulfsohn1997}.  In the joint modeling approach, the longitudinal measurement is assumed to follow a linear mixed model with normal measurement error \citep{laird1982}, and the failure time is modeled through the proportional hazards model. The time-dependent covariate is taken as the unobserved longitudinal process \citep{tsiatis2001}. Statistical inference is carried out by likelihood or conditional likelihood. {Furthermore, \cite{song2008} and \cite{Andrinopoulou2018} 
extended the constant hazard ratio in classic Cox model to a time-varying one.} A recent review of the joint modeling approach can be found in \cite{rizopoulos2012}. As a likelihood-based method, joint modeling imposes rather strong modeling assumptions, and inferences are quite complicated \citep{Song2002, Rizopoulos2009}. Furthermore, if the model is misspecified, bias occurs, and standard deviations may not be computable \citep{cao2015a, arisido2019}. 
Despite the fact that joint modeling imposes strong modeling assumptions and has complicated computation and inference, it allows 
summary trends, such as slope or spread, to enter the survival model as covariates. Such features make the model more coherent and interpretable. 

In this paper, we propose a varying coefficient model for censored outcomes with intermittently observed time-dependent covariates. The hazard function is specified as follows: 
 \begin{equation}\label{equ:tcox}
\lambda\{t\mid Z(r), r \leq t\}=\lambda_0(t)e^{\beta_0(t)^{\mathrm{T}}Z(t)},
\end{equation}
where $\lambda_0(t)$ is the baseline hazard function, $Z(t)$ is the longitudinally collected time-dependent covariates and $\beta_0(t)$ is a non-parametric function describing the multiplicative relationship between $Z(t)$ and the hazard function. This flexible modeling framework allows the hazard ratio to change over time, which is more informative and realistic for many practical situations. A na\"ive approach that imputes $Z(t)$ at failure time by the most recent longitudinal observation and implements the method in \cite{tian2005} results in biased coefficient estimation. The bias does not attenuate with increased sample size, as demonstrated in our simulation studies. We propose an estimating equation-based weighting strategy without imposing stringent distributional assumptions on the longitudinal process. To estimate $\beta_0(t)$ in (\ref{equ:tcox}), we use a two-dimensional kernel function to do smoothing. At any fixed time point $t,$ we estimate $\beta_0(t)$ by solving an estimating equation, where higher weights are given to the longitudinal observations that have measurement times close to the failure time, and lower weights are given to those that have measurement times far from the failure time. 
Unlike the last value carried forward method, which imputes the most recent observation regardless of the distance between its measurement time and the failure time, by adaptive weighting, we get an asymptotically unbiased coefficient estimation. With time-invariant coefficient $\beta_0$, \cite{cao2015a, cao2021} studied the proportional hazards model (\ref{equ:cox}) for sparse longitudinal covariates. 

Moreover, we construct simultaneous confidence bands (SCBs) for $\beta_0(t).$ Specifically, for a pre-specified confidence level $1-\alpha,$ we aim to find random functions $L(t)$ and $U(t)$ such that
$$
P\{L(t)\leq \beta_0(t) \leq U(t), t \in[b_1,b_2]\} \rightarrow 1-\alpha
$$
as the number of subjects $n \rightarrow \infty$ in some closed interval $[b_1,b_2].$ Unlike point-wise confidence intervals, a simultaneous confidence band covers the underlying non-parametric function with a pre-specified probability. For a non-parametric function, it is more informative to evaluate the overall pattern and magnitude of the variation. In addition, a confidence band is often graphically intuitive and versatile, especially when investigators do not know {\it a priori} the precise hypothesis of interest. 

 To construct SCBs, one has to derive the asymptotic distribution of the maximum deviations between the estimated and the true coefficient functions. 
The convergence of such asymptotics is slow, of $\log n$ rate \citep{cao2018}. In practice, a multiplier bootstrap, where standard normal variates are introduced into an appropriate statistic, keeping the data fixed, is usually adopted. The distribution of the limiting process is approximated via a large number of realizations, repeatedly generating standard normal variates \citep{lin1993, lin1997}. Later developments include using Poisson multipliers and other zero mean and unit variance multipliers 
\citep{tian2005, beyersmann2013, dobler2014, dobler2019}.  

The rest of the paper is organized as follows. In Section 2, we use a kernel-weighted estimating equation to estimate the non-parametric function $\beta_0(t)$ in (\ref{equ:tcox}). We establish its asymptotic normality for any fixed time point. Furthermore, we construct a simultaneous confidence band to evaluate the overall magnitude of variation. 
A series of simulation studies in Section 3 illustrate that the proposed method works well in finite samples and has improved performance over the last value carried forward approach {and joint modeling approach}. 
Section 4 applies the proposed method to analyze {a dataset from the Alzheimer's Disease Neuroimaging Initiative study. }
 Concluding remarks are given in Section 5. 
All proofs are relegated to the Supplementary Material. 

\section{Estimation and Inference}\label{2.sec}
\subsection{Problem Set Up and Notations}
Suppose that we have a random sample of $n$ independent subjects. For the $i$-th subject, let $T_i$ denote the failure time, and $C_i$ denote the 
censoring time. It is assumed that
censoring is coarsened at random such that $T_i$ and $C_i$ are independent given the covariate process $Z_i(\cdot)$  \citep{heitjan1991}. Denote $X_i=\min(T_i, C_i)$ and $\delta_i = I(T_i \le C_i).$ The $p$-dimensional covariate process $Z_i(\cdot)$ may include both time-independent and time-varying covariates. It is assumed that 
the time-varying covariates are observed at the same time points within individuals.  The longitudinal covariates are observed at $M_i$
observation times $R_{ik}, k = 1,\ldots, M_i,$ where $M_i$ is assumed finite with probability one. 
The observed data consist of $n$ 
independent realizations of $\left(X_i,\delta_i, R_{ik}, Z_i(R_{ik}), k = 1,..., M_i\right),$ $i = 1, ..., n.$ 
We use the counting process to denote $N_i(t)=I(X_i \leq t,\delta_i=1)$, $Y_i(t)=I(X_i \geq t),$ and $N_i^*(t) = \sum_{k=1}^{M_i} I\{R_{ik} \leq t\}, i =1, \ldots, n.$

\subsection{Estimation}
We first write the partial likelihood with a time-varying coefficient function as follows. 
\begin{equation*}
L_n\{\beta(t),t\}=\prod_{i=1}^n \left\{ \frac{e^{\beta(t)^TZ_i(t)}}{\sum_{j=1}^n Y_j(t)e^{\beta(t)^TZ_j(t)}}\right\}^{\Delta N_i(t)},
\end{equation*}
where
$$
\Delta N_{i}(t)= \begin{cases}1 & \text { if } N_{i}(t)-N_{i}(t-)=1 \\ 0 & \text { otherwise. }\end{cases}
$$

The log partial likelihood function is
\begin{equation}\label{oracle}
l_n\{\beta(t),t\}=n^{-1}\sum_{i=1}^n \int_{0}^{\tau}\left[\beta(t)^T  Z_i(t)-\log\{\sum_{j=1}^n Y_j (t)e^{\beta(t)^T Z_j(t)}\}\right]dN_i(t),
\end{equation}
{ where $\tau$ is the pre-specified maximum observation time.}
With a time-independent coefficient $\beta,$ \cite{cao2015a} proposed a kernel smoothing approach to handle the sparse longitudinal covariates. When the entire trajectory of $Z(t)$ is assumed to be known, \cite{tian2005} proposed to estimate the time-varying effect $\beta_0(t)$ in (\ref{equ:tcox}) through kernel smoothing.
To simultaneously incorporate both the time-varying effect and sparsely observed longitudinal covariates, we use a bivariate kernel function to perform the smoothing. Specifically, we propose the following estimating equation 
\begin{equation}\label{coxestequation}
U_n\{\beta(s)\}=n^{-1}\sum_{i=1}^n\sum_{k=1}^{M_i}\int_0^{\tau}K_{h_1,h_2}(t-s,R_{ik}-s)[Z_i(R_{ik})-\bar{Z}\{\beta(s),t\}] d N_i(t),
\end{equation}
where $K_{h_1,h_2}(t,s)=K(t/h_1,s/h_2)/(h_1h_2),$ $K(\cdot,\cdot)$ is a bivariate kernel function, 
and
\begin{align*}
 &\bar {Z}\{\beta(s),t\}=\frac{S^{(1)}\{\beta(s),t\}}{S^{(0)}\{\beta(s), t\}},  \\
 &S^{(l)}\{\beta(s),t\}=\frac{1}{n}\sum_{j=1}^n \sum_{k=1}^{M_j} K_{h_1,h_2}(t-s,R_{jk}-s)Y_j(t)Z_j(R_{jk})^{\otimes l}\exp\{\beta(s)^T Z_j(R_{jk})\}, 
  l=0, 1, 2,
  \end{align*}
where $a^{\otimes 0}=1, a^{\otimes 1}=a,$ and $a^{\otimes 2}=aa^T.$ 

Denote 
\begin{equation}\label{longitudinal2}
\E[dN^*(t) ] = \lambda^*(t) d t,
\end{equation}
where $\lambda^*(t)$ is positive and twice continuously differentiable for all $t \in [0, \tau].$ Define $$ s^{(l)}\{\beta(t),t\}=\E\left[ Y(t)Z(t)^{\otimes l}\exp\left\{\beta(t)^T Z(t)\right\}\right]\lambda^*(t), $$
as the limit of $ S^{(l)}\{\beta(t),t\}, l=0,1,2.$
For any fixed time point $s \in [0, \tau],$ we weigh the contribution from the longitudinally observed covariate process and the failure time by their distances to $s$ using two bandwidths. Solving the estimating equation (\ref{coxestequation}), we obtain $\hat{\beta}(s)$ as the estimation of the non-parametric regression function ${\beta}_0(s)$ in (\ref{equ:tcox}). In practice, we use a quasi-Newton method \citep{broyden1965} to solve the estimating equation (\ref{coxestequation}). To state its asymptotic properties, we need the following conditions.


\begin{enumerate}[label=(A\arabic*)]
\item  \label{con:count} 
(\ref{longitudinal2}) holds. 
{The longitudinal observation process $N_i^*(\cdot)$ is independent of the observed longitudinal covariate $Z_i(R_{ik}), i=1, \ldots, n;$ $k = 1, \ldots, M_i.$ In addition, censoring time is non-informative in the sense that $C_i$ is independent of longitudinal observation time $R_{ik}$ and the observed longitudinal covariate $Z_i(R_{ik})$ in addition to that $T_i$ and $C_i$ are independent given the covariate process $Z_i(\cdot), i =1, \ldots, n; k = 1, \ldots, M_i.$}
Moreover, with the pre-specified constant $\tau,$ $N^*(\tau)$ is bounded by a finite constant. 
{Additionally, we require $\tau$ to satisfy $P(X \geq \tau)>0.$} 

\item  \label{con:var}  For any fixed time point $t\in[0,\tau],$ $B\{\beta_0(t),t\}$ is non-singular, where

 $$
 B\big\{\beta_0(t),t\big\} = \left[{s}^{(2)}\{\beta_0(t),t\}-\frac{{s}^{(1)}\{\beta_0(t),t\}^{\otimes 2}}{{s}^{(0)}\{\beta_0(t),t\}}\right]\lambda_0(t) .
 $$
\item \label{con:continuous}  For any fixed time point $s\in[0,\tau],$ $\E\left[ Z(s+t_2)
Y(s+t_1)e^{\beta_0(s+t_1)^TZ(s+t_1)}\right]$ and \newline $\E\left[\bar{Z}\{\beta_0(s),s+t_1\}
Y(s+t_1)e^{\beta_0(s+t_1)^TZ(s+t_1)}\right]$ are twice continuously differentiable for \newline 
$(t_1, t_2) \in [0, \tau]^{\otimes 2}$ and $(s+t_1, s+t_2) \in [0, \tau]^{\otimes 2}.$ 
\item \label{con:kernel} $K(x,y)$ is a symmetric bivariate density function. In addition, $\iint x^2 K(x,y)d x d y < \infty,$  $\iint y^2 K(x,y)d x d y < \infty,$  and $\int K(x,y)^2 d x d y < \infty.$ Moreover, as $n \rightarrow \infty,$ $nh_1h_2\rightarrow \infty$ and $(nh_1h_2)^{1/2}(h_1^2+h_2^2) \rightarrow 0.$  
\item \label{con:covariate}  
The covariate process $Z(t)$ has bounded total variation on  $[0,\tau]$ almost surely.   
\end{enumerate}

Condition \ref{con:count} assumes that the observational time 
{and the censoring time} 
is noninformative. Analogous assumptions have been 
assumed in \citep{cao2021, sun2022}. 
{The assumption $P(X\geq \tau)>0$ guarantees that $E\left[Y(t)\exp\{\beta_0(t)^TZ(t)\}\right]\lambda^*(t) >0, \forall t\in [0,\tau]$, which implies that the limit of the denominator of $\bar{Z}\{\beta_0(t),t\}$ in equation (\ref{coxestequation}) is bounded away from $0$ on $[0,\tau]$ \citep{andersen1982}.} 
  Condition \ref{con:var} ensures the identifiability of $\beta_0(t)$ at any fixed time point $t\in[0,\tau].$ 
Condition \ref{con:continuous} posits smoothness assumptions on the expectation of certain functions of the covariate process. Condition \ref{con:kernel} specifies valid kernels and bandwidths. Condition \ref{con:covariate}  is common for time-dependent covariates. 

The asymptotic property of the non-parametric function 
$\hat{\beta}(\cdot)$ is detailed in the following theorem:

\begin{theorem}\label{th:coxnormal}
 Under conditions \ref{con:count}-\ref{con:covariate}, for any fixed time point $s \in [h,\tau-h],$ where $h= h_1 \lor h_2$ is a small positive number, 
 the asymptotic distribution of $\hat{\beta}(s)$ satisfies
 $$
(nh_1h_2)^{1/2}B\big\{\beta_0(s),s\big\}\{\hat{\beta}(s)-\beta_0(s)\}
 \stackrel{d}{\rightarrow} N\big(0,\Sigma(\beta_0(s),s)\big)
 $$
 where
 $$
 \Sigma(\beta_0(s),s)=\iint  K(z_1,z_2)^{2}   \left\{{s}^{(2)}\left(\beta_{0}(s), s\right)-\frac{{s}^{(1)}\left(\beta_{0}(s), s\right)^{\otimes 2}}{{s}^{(0)}\left(\beta_{0}(s), s\right)}\right\}\lambda_0(s)d z_1 d z_2. 
 $$
 \end{theorem}
In practice, we use the estimating equation (\ref{coxestequation}) and estimate $\Sigma(\beta_0(s),s)$ by 
 $$
  \hat{\Sigma}(\hat{\beta}(s),s)=n^{-2}\sum_{i=1}^n \left( \sum_{k=1}^{M_i}\int_0^{\tau}K_{h_1,h_2}(u-s,R_{ik}-s)\left[
 Z_i(R_{ik})-\bar{Z}\{\hat{\beta}(s),u\}\right]d N_i(u)\right)^{\otimes2}.
 $$
The variance of $\hat{\beta}(s)$ can be estimated by the sandwich formula
\begin{align}\label{sandwich}
\Bigg(\frac{\partial  U_n\{\beta(s)\} }{\partial\beta(s)}\bigg|_{\beta(s)=\hat{\beta}(s)}\Bigg)^{-1} \hat{\Sigma}(\hat{\beta}(s),s)\Bigg(\frac{\partial  U_n\{\beta(s)\} }{\partial\beta(s)}\bigg|_{\beta(s)=\hat{\beta}(s)}\Bigg)^{-1}.
\end{align}

{
In the proof presented in the Supplementary Material, the asymptotic bias is of order $(nh_1h_2)^{1/2}O_p(h_1^2+h_1h_2+h_2^2),$ which vanishes under \ref{con:kernel}.
}
Under \ref{con:kernel}, 
{ the proposed estimator} $\hat{\beta}(s)$ is consistent for 
any $s\in[h,\tau-h].$  
\begin{corollary} \label{cor1}
Under conditions \ref{con:count}-\ref{con:covariate}, the sandwich formula (\ref{sandwich}) consistently estimates the variance of $\hat{\beta}(s),$  for any fixed time point $s \in [h,\tau-h],$ where $h= h_1 \lor h_2$ is a small positive number.
\end{corollary}

\subsection{The Construction of Simultaneous Confidence Band}
For the unknown function $\beta_0(\cdot),$ it is more informative to construct a simultaneous confidence band in addition to pointwise inference in Theorem \ref{th:coxnormal}. Specifically, for a pre-specified $\alpha,$ and a smooth function $l(t)\in R^p,$ 
we aim to find smooth random functions $L(t)$ and $U(t)$ that satisfy
$$
P\left\{L(t) \le l(t)^T\beta_0(t) \le U(t), \forall t \in [h, \tau-h]\right\} \rightarrow 1-\alpha
$$
as number of subjects $n \rightarrow \infty.$ 
To obtain such SCBs for $l(t)^T\beta_0(t),t \in 
[h,\tau-h]$, 
we need to derive a large-sample  approximation to the distribution of
\begin{equation} \label{scb}
{\mathcal{S}_{SCB}}=\sup _{t \in[h,\tau-h]}  \hat{w}(t)  \left|l(t)^T\left\{ \hat{\beta}(t)-\beta_0(t) \right\}\right|,
\end{equation}
where $\hat{w}(t)$ is a possibly data-dependent, positive weight function. It converges uniformly to a deterministic function. For instance, when $\beta_0(t)$ is a scalar, we can define $\{\hat{w}(t)\}^{-1}$ as the estimated standard error of $\hat{\beta}(t)$. Such weighting prevents the domination of time points with large variances, which may lead to 
unnecessarily wider confidence band.


Inspired by \cite{tian2005} and \cite{dobler2019},  we consider a stochastic perturbation of the estimating equation. 
 \begin{equation} \label{perturbation-cox}
\tilde{U}_n\{ {\beta}(s)\}= n^{-1}\sum_{i=1}^n\sum_{k=1}^{M_i}\int_0^{\tau}K_{h_1,h_2}(t-s,R_{ik}-s)\left[Z_i(R_{ik})-\bar{Z}\{{\beta}(s),t\}\right]\xi_{i} d N_i(t),
\end{equation}
 where  $ \xi_{i}\ (i=1,\ldots,n)$ are i.i.d. random variables with zero mean and unit variance and are independent of the data $\{(X_i,\delta_i, Z_i(R_{ik}), R_{ik}, k = 1,\ldots, M_i), i = 1, ..., n\}$. For example, $\xi_i$ can be $N(0,1)$ or the Rademacher variable, which takes values $+1$ and $-1$ with equal probability. 
Then, conditional on the data $\{(X_i,\delta_i, Z_i(R_{ik}), R_{ik}, k = 1,\ldots, M_i), i = 1, ..., n\}$, the distribution of
\begin{align}\label{tilderscb}
  \tilde{\mathcal{S}}_{SCB}   =\sup _{s \in[h,\tau-h]} \hat{w}(t)\left|l(s)^TI\{\hat{\beta}(s)\}^{-1} \tilde{U}_n\{ \hat{\beta}(s)\}\right|
\end{align}
 can be used to approximate the unconditional distribution of 
 $\mathcal{S}_{SCB}$ in (\ref{scb}), where
\begin{align*}
 &I\{\hat{\beta}(s)\} = \frac{-1}{n} \sum_{i=1}^n\sum_{k=1}^{M_i}\int_0^\tau K_{h_1,h_2}(t-s,R_{ik}-s)Y_i(t)\left[\frac{S^{(2)}\{\hat{\beta}(s),t\}}{S^{(0)}\{\hat{\beta}(s),t\}}-\left(\frac{S^{(1)}\{\hat{\beta}(s),t\}}{S^{(0)}\{\hat{\beta}(s),t\}}\right)^{\otimes 2}\right]  d t.
 \end{align*}
 We repeat this process $B$ times to obtain $\left \{ \tilde{\mathcal{S}}_{SCB}^{(1)}, \ldots,  \tilde{\mathcal{S}}_{SCB}^{(B)} \right \}$. For instance, we can take $B = 5,000.$ Denote its $1-\alpha$ empirical percentile as $c_\alpha.$ The SCB for $l(t)^T\beta_0(t)$ can be written as 
 \begin{equation}
l(t)^T \hat{\beta}(t) \pm c_\alpha \{\hat{w}(t)\}^{-1}. \label{equ:scb}
 \end{equation}
This is computationally efficient as we do not need to find the non-linear function $\hat{\beta}(t)$ $B$ times. We only need to evaluate $\tilde{U}_n\{\cdot\}$ 
at $\hat{\beta}(t)$ $B$ times. 
In practice, for instance, if we are interested in the regression coefficient function of the first covariate, we can take $l(t) = (1, \ldots, 0).$ We provide a theoretical justification of this procedure in the Supplementary Material. Numerical studies show that this procedure works well with a moderate sample size, and the nominal coverage can be obtained.

\subsection{Bandwidth Selection}\label{sec2.4}
Choosing a suitable bandwidth is practically important. We propose to estimate the bias and variability separately and choose the bandwidth that minimizes the mean squared error \citep{cao2015b}. 
For a fixed time point $t\in[h,\tau-h],$ 
we regress $\hat{\beta}(h_1,h_2,t)$ on $b=(h_1^2, h_1h_2, h_2^2)^T $ in a reasonable range of the bandwidths to obtain the slope estimate $\hat{C} = (c_1, c_2, c_3)^T$ for bias term calculation. For the variance term, we split the data randomly into two parts and obtain regression coefficient estimates $\hat{\beta}_{1}(h_1,h_2,t)$ and $\hat{\beta}_{2}(h_1,h_2,t)$ based on each part. The variance of $\hat{\beta}(h_1,h_2,t)$ is then estimated by $\hat{V}(h_1,h_2,t)=\left\{\hat{\beta}_{1}(h_1,h_2,t)-\hat{\beta}_{2}(h_1,h_2,t)\right\}^{2} / 4$. Using both $\hat{C}$ and $\hat{V}(h_1,h_2,t)$, we thus calculate the 
mean-squared error as $(\hat{C}^{T} b)^2+\hat{V}(h_1,h_2,t)$ for $t.$ 
We repeat the above procedure 
at equally spaced time points and sum them up to obtain integrated mean-square errors. The optimal bandwidth is the one that minimizes this summation.

\section{Numerical Studies}\label{4.sec}
In this section, we evaluate finite sample performance of the proposed method through simulations. We repeatedly generate 
$1,000$ datasets with sample sizes 400 and 900. The end of follow-up time $\tau$ is specified as 1.

\subsection{Data Generating Process}
We first generate the time dependent covariate process $Z(t)$ 
{based on a} Gaussian 
{ process} with mean $-1-2(t-1)^2$ and variance covariance matrix $\mbox{Cov}\{Z(t), Z(s) \} = e^{-|t-s|},$ 
{where $t, s\in (0,1).$ } 
{Specifically, 
we generate the covariate process through the piecewise constant function
$$
Z(t)=\sum_{i=1}^{20}I\{(i-1)/20 \leq t < i/20 \}z_i,
$$ 
where $(z_i)_{i=1}^{20}$ follows a multivariate normal distribution with mean $-1-2((i-1)/20-1)^2$ and variance $1.$ The covariance between $z_i$ and $z_j$ is $e^{-|i-j|/20}.$ }
The number of observations is generated from $\mbox{Pois}(5)+1$, and the observational times are generated from $\mathcal{U}(0,1),$ where $\mbox{Pois}(5)$ means a Poisson distributed random variable with mean $5$ and $\mathcal{U}(0,1)$ means standard uniform distribution. 

The failure time $T$ is generated from the multiplicative hazards model with 
varying-coefficient 
as follows.
\begin{align*}
\lambda\left\{t \mid Z(s),s\le t\right\}=\lambda_0(t)\exp\left\{\beta_0(t)^T Z(t) \right\},
\end{align*}
where $\lambda_0(t)=2+0.1t$ and $\beta_0(t)=0.5\sin(2\pi t)$.
{To generate the failure time, we first generate a random variable $u$ from $\mathcal{U}(0,1).$ Then, we solve $S(t)=u,$ where the survival function takes the form $$S(t)=\exp\{-\int_0^t \lambda\{v|Z(u),u\leq v\} dv\}=\exp\{-\int_0^t \lambda_0(v)\exp\{\beta_0(v)^TZ(v)\} dv\}.$$
To approximate the integral in $S(t)$, we use Gauss Legendre quadrature \citep{Abramowitz1983}, which is 
implemented using package $\texttt{gaussquad}$ in R.}
The censoring time is generated from $\min\left(1,C^{*}\right),$ where $C^{*}\sim \mathcal{U}\left(\gamma,1.5\right)$ giving censoring percentages of $15\%$ and $35\%,$ respectively, by changing $\gamma$. 
{For each fixed time point $t$, we construct confidence intervals by solving for (\ref{coxestequation}) to obtain an estimate of $\beta_0(t)$ and an estimate of variance with sandwich form 
 (\ref{sandwich})}

\subsection{Comparison with Last Value Carried Forward  and Joint Modeling Approach}
With time-dependent covariate, if the longitudinal covariate is unavailable at the failure time, the most recently observed longitudinal covariate is used instead. Such imputation is intuitively appealing yet ignores the dynamics of the longitudinal process yielding biased results. {Joint modeling is the most widely used method to simultaneously handle longitudinal
covariates and censored outcome. Despite its popularity, the modeling assumptions are quite strong with complicated inferences and computations. 
In this subsection, we compare the {\it ad-hoc} last value carried forward method and the  commonly used joint modeling approach with the proposed kernel weighting approach.
}



First, we compare our method with the last value carried forward method. We report the results with $h_1=n^{-0.35}$ and $h_2=n^{-0.35}$ or $ n^{-0.45}$, since those bandwidths 
give stable results. Results based on the automatic bandwidth selection rule are also provided.
We summarize the results based on the proposed method 
for different sample sizes and censoring rates in Table \ref{table1}, 
where ``auto" refers to bandwidths determined using the adaptive selection technique described in Section \ref{sec2.4}. The biases are small at different time points. The $\hat{\beta}(t)$ variance estimator is accurate.
The coverage probabilities are close to the nominal one. 
Estimation based on data adaptive bandwidth selection has satisfactory performance. The performance of estimators under varied censoring rates does not differ much. 
The performance improves with increased sample size.

\begin{table}[htbp]
\caption{Simulation results of $\hat{\beta}(s)$. } 
\begin{center}\small
\label{table1}
\begin{tabular}{llccrccccrccc}
\toprule
  & &  &   & \multicolumn{4}{c}{Censoring\ rate\ is\ $15\%$} &  & \multicolumn{4}{c}{Censoring\ rate\ is\ $35\%$}\\
\cline{5-8}  \cline{10-13}
$s$ & $n$ & $h_1$ & $h_2$ & Bias & SE & SD & CP &  & Bias & SE & SD & CP\\
\midrule
\multicolumn{12}{c}{Proposed method}\\
$0.2$   & 400 & $n^{-0.35}$ & $n^{-0.35}$& -0.072  & 0.171 & 0.169 & 91.6 &  &  -0.073   & 0.177 & 0.174 & 91.9\\
        &     & $n^{-0.35}$ & $n^{-0.45}$ & -0.053   & 0.207 & 0.203 & 93.4 &  &  -0.053  & 0.217 & 0.209 & 92.9\\
        &     & auto      &       &  -0.058    & 0.198 & 0.193 & 93.3 &  &  -0.060  & 0.207 & 0.197 & 92.2\\
        & 900 & $n^{-0.35}$ & $n^{-0.35}$ & -0.042      & 0.150 & 0.146 & 93.3 &  & -0.044     & 0.153 & 0.150 & 92.7\\
       &     & $n^{-0.35}$ & $n^{-0.45}$ & -0.030    & 0.187 & 0.181 & 93.2 &  & -0.031   & 0.193 & 0.186 & 92.2\\
          &     & auto    &          & -0.031  & 0.178 & 0.171 & 93.3 &  & -0.033  & 0.182  & 0.174  & 92.5 \\
$0.4$   & 400 & $n^{-0.35}$ & $n^{-0.35}$& -0.056   & 0.156 & 0.153 & 92.0 &  &  -0.053     & 0.189 & 0.176 & 90.6\\
        &     & $n^{-0.35}$ & $n^{-0.45}$ & -0.052    & 0.194 & 0.183 & 91.3 &  &  -0.046  & 0.232 & 0.210 & 90.8\\
 &     & auto      &        &  -0.055   & 0.190 & 0.178 & 90.2 &  &  -0.049  & 0.216 & 0.199 & 90.3\\
        & 900 & $n^{-0.35}$ & $n^{-0.35}$ & -0.042       & 0.139 & 0.130 & 91.2 &  & -0.040    & 0.156 & 0.147 & 91.5\\
        &     & $n^{-0.35}$ & $n^{-0.45}$& -0.040     & 0.173 & 0.162 & 92.0 &  & -0.038    & 0.195 & 0.183 & 92.2\\
           &     & auto      &        &  -0.040  & 0.164 & 0.154 & 91.6 &  &  -0.038  & 0.184 & 0.171 & 92.2\\
$0.6$   & 400 & $n^{-0.35}$ & $n^{-0.35}$&  0.039  & 0.133 & 0.129 & 93.0 &  &   0.044   & 0.166 & 0.163 & 92.9\\
         &     & $n^{-0.35}$ & $n^{-0.45}$ &  0.035        & 0.160 & 0.153 & 92.4 &  &  0.042     & 0.198 & 0.194 & 93.0\\
      &     & auto     &         &  0.036   & 0.152 & 0.147 & 93.1 &  &  0.041  & 0.187 & 0.184 & 93.0\\
        & 900 & $n^{-0.35}$ & $n^{-0.35}$&   0.027      & 0.106 & 0.105 & 93.8 &  &  0.029     & 0.134 & 0.133 & 92.7\\
        &     & $n^{-0.35}$ & $n^{-0.45}$&  0.021     & 0.134 & 0.131 & 93.3 &  &   0.023   & 0.167 & 0.165 & 92.6\\
           &     & auto    &          &  0.021  & 0.126 & 0.124 & 93.8 &  &  0.025  & 0.158 & 0.154 & 92.1\\
$0.8$   & 400 & $n^{-0.35}$ & $n^{-0.35}$&  0.038     & 0.201 & 0.189 & 91.3 &  &  0.022     & 0.285 & 0.250 & 90.0\\

       &     & $n^{-0.35}$ & $n^{-0.45}$ &  0.024     & 0.241 & 0.225 & 92.4 &  &   -0.006    & 0.347 & 0.307 & 90.8\\
      &     & auto    &          &  0.027   & 0.231 & 0.214 & 91.5 &  &  0.006  & 0.322 & 0.282 & 90.1\\
        & 900 & $n^{-0.35}$ & $n^{-0.35}$&  0.031    & 0.166 & 0.158 & 92.1 &  &  0.013        & 0.229 & 0.208 & 92.7\\
        &     & $n^{-0.35}$ & $n^{-0.45}$&  0.018    & 0.209 & 0.197 & 92.6 &  &   -0.010      & 0.287 & 0.260 & 91.5\\
     &     & auto      &        &  0.021  & 0.197 & 0.186 & 92.1 &  &  -0.003  & 0.269 & 0.243 & 91.5\\

 \multicolumn{13}{c}{LVCF}\\
  $0.2$  &  400 & $n^{-0.35}$ && -0.094 & 0.167 & 0.160 & 89.0 && -0.096 & 0.174 & 0.164 & 88.4\\
         &  900 & $n^{-0.35}$ && -0.077 & 0.128 & 0.125 & 89.7 && -0.078 & 0.133 & 0.129 & 88.9\\
  $0.4$  &  400 & $n^{-0.35}$ && -0.082 & 0.127 & 0.123 & 88.2 && -0.076 & 0.143 & 0.139 & 90.1\\
         &  900 & $n^{-0.35}$ && -0.071 & 0.093 & 0.095 & 88.5 && -0.069 & 0.107 & 0.107 & 89.8\\
  $0.6$  &  400 & $n^{-0.35}$ &&  0.088 & 0.095 & 0.095 & 83.7 &&  0.096 & 0.119 & 0.119 & 85.4\\
         &  900 & $n^{-0.35}$ &&  0.080 & 0.069 & 0.071 & 78.8 &&  0.084 & 0.088 & 0.089 & 83.0\\
  $0.8$  &  400 & $n^{-0.35}$ &&  0.112 & 0.139 & 0.134 & 83.5 &&  0.100 & 0.192 & 0.180 & 87.3\\
         & 900  & $n^{-0.35}$ &&  0.115 & 0.106 & 0.103 & 76.1 &&  0.109 & 0.144 & 0.138 & 83.7\\
\bottomrule
\end{tabular}
\end{center}
\footnotesize{
Note: ``BD'' represents different bandwidths, ``Bias'' is the difference between $\beta_0(s)$ and $\hat{\beta}(s)$,
``SD'' is the sample standard deviation, ``SE'' is the average of the standard error estimates, ``CP''$/100$ represents the coverage probability of the $95\%$ confidence interval for estimators of $\beta_0(s)$ at fixed $s$, and LVCF represents the last value carried forward method. 
}
\end{table}
Table \ref{table1} also shows results based on the  
last value carried forward.
We observe that bias occurs, which does not attenuate with increased sample size. As a result, the coverage probabilities are lower than the nominal ones. Paradoxically, the coverage probabilities get worse with a decreased censoring rate. The reason is that with lower censoring rate, more data are used for parameter estimation, producing a larger bias. 
{We use covariates observed before the minimum of censoring and death time.  When the observation process follows a homogeneous Poisson process, as time $s$ increases, the amount of observation decreases, and the information used to estimate $\beta(s)$ also decreases. Consequently, the standard deviation becomes larger. It is a coincidence that the bias changes with 
$s$. In other settings, such as the ones summarized in the Supplementary Material, the bias does not decrease with time.}

Next, we compare the proposed method with joint modeling approach. Specifically, we employed the joint modeling approach proposed by \cite{Andrinopoulou2018}, which utilizes P-splines to estimate the time-varying coefficient function. P-splines extend B-splines by incorporating a penalty term to regulate smoothness and mitigate overfitting \citep{eilers1996}. 
The simulation setup follows the same configuration as above. In this scenario, the censoring rate is $15\%$. We use automatic bandwidth selection for our method.

Due to the extensive computational time required by the joint modeling approach, we present results based on 100 replications. Additionally, since \cite{Andrinopoulou2018} does not explicitly state the estimation of 
the time-varying coefficients $\beta_0(\cdot)$, its standard deviation or the construction of confidence intervals, we follow the approach described in \cite{rizopoulos2012}. Specifically,  we use the median of the post-burn-in Monte Carlo samples to estimate 
$\beta(s)$, the standard deviation of the post-burn-in Monte Carlo samples to estimate the standard deviation of $\hat{\beta}(s)$ (``SE" in Table \ref{tableJM}) , and the 0.025 and 0.975 quantiles of the samples to construct the $95\%$ confidence intervals. The results based on the mean of the post-burn-in Monte Carlo samples are similar and thus omitted. 
The simulation results are summarized in Table \ref{tableJM}.
 \begin{table}[htbp]
\caption{Simulation results of $\hat{\beta}(s)$. } 
\begin{center}\small
\label{tableJM}
\begin{tabular}{ccrccccrccc}
\toprule
   &   & \multicolumn{4}{c}{Joint modeling approach} &  & \multicolumn{4}{c}{The proposed method}\\
\cline{3-6}  \cline{8-11}
$n$ & $s$ & Bias & SD & SE & CP &  & Bias & SD & SE & CP \\
\midrule 
100   & $0.2$ & -0.439  & 0.319 & 0.276  & 64.0& &  -0.123   & 0.233 & 0.213 & 86.0\\
    & $0.4$ & -0.225   & 0.278 & 0.255  & 83.0& &  -0.113     & 0.214 & 0.221 & 89.0\\
    & $0.6$ & 0.304  & 0.362 & 0.305 & 79.0& &   0.033   & 0.192 & 0.189 & 93.0\\
   & $0.8$ &  0.417     & 0.525 & 0.424  & 72.0& &  0.055     & 0.239 & 0.260 & 96.0\\
200   & $0.2$ & -0.456  & 0.265 & 0.270  & 62.0& &  -0.062   & 0.231 & 0.210 & 92.0\\
    & $0.4$ & -0.234   & 0.276 & 0.257 & 85.0& &  -0.030     & 0.191 & 0.194 & 94.0\\
    & $0.6$ & 0.366  & 0.430 & 0.315 & 71.0 & &   0.035   & 0.152 & 0.166 & 96.0\\
   & $0.8$ &  0.517    & 0.639 & 0.432  & 68.0& &  0.027     & 0.254 & 0.225 & 91.0\\
\bottomrule
\end{tabular}
\end{center}
\footnotesize{
Note: `Bias'' is the difference between $\beta_0(s)$ and $\hat{\beta}(s)$,
``SD'' is the sample standard deviation, ``SE'' is the average of the standard deviation estimates, ``CP''$/100$ represents the coverage probability of the $95\%$ confidence interval for $\hat{\beta}(s).$ 
}
\end{table}
As shown in Table \ref{tableJM}, the estimates obtained using the joint modeling approach exhibit substantial biases resulting in poor coverage probabilities. 
In comparison, the proposed method has decent performances.
The bias is small, and decreases with increased sample size.
Additionally, SE agrees with SD, and the coverage probabilities align well with the 
nominal level of $95\%$.
Furthermore, the joint modeling approach requires a large number of iterations when employing Markov chain Monte Carlo method, making it computationally expensive.
For instance, with a sample size of 100,  using a computer equipped with an Intel Core i7-12700H CPU (4.7 GHz) and 64 GB of RAM,  a single run of the joint modeling approach takes approximately 45 minutes. In contrast, the proposed method 
finishes the computation within a few seconds.


\subsection{Constructing Simultaneous Confidence Band}
We are interested in constructing SCBs for $\beta_0(s)$ in this subsection. We use 50 equally spaced grid points in $[h,1-h]$ to calculate coverage probability. Here, all intervals are constructed based on $M=5000$ realizations of $\{\xi_i \stackrel{i.i.d.}{\sim} \mbox{Exp}(1)-1, i=1,\ldots,n\},$ where $\mbox{Exp}(1)$ represents the exponential distribution with mean 1. We tried other $\xi_i, i =1, \ldots, n$, and the results are similar, and thus omitted. 
{In the simulation study and the real data analysis, we use the inverse of the estimated standard error of $\hat{\beta}(s)$ as the weight $\hat{w}(s).$  In this case, the weighted estimation bias $\hat{w}(s)\left\{\hat{\beta}(s)-\beta_0(s)\right\}$ has the same distribution at all time points. Then, the contribution to the simultaneous confidence band is the same at different time points, even if the variance of the estimator is different. Such a choice can avoid the influence of points with larger variance dominating the influence of other points and can lead to a narrower band. }

\begin{figure}[htb!]
\begin{center}
\includegraphics[width=0.49\linewidth]{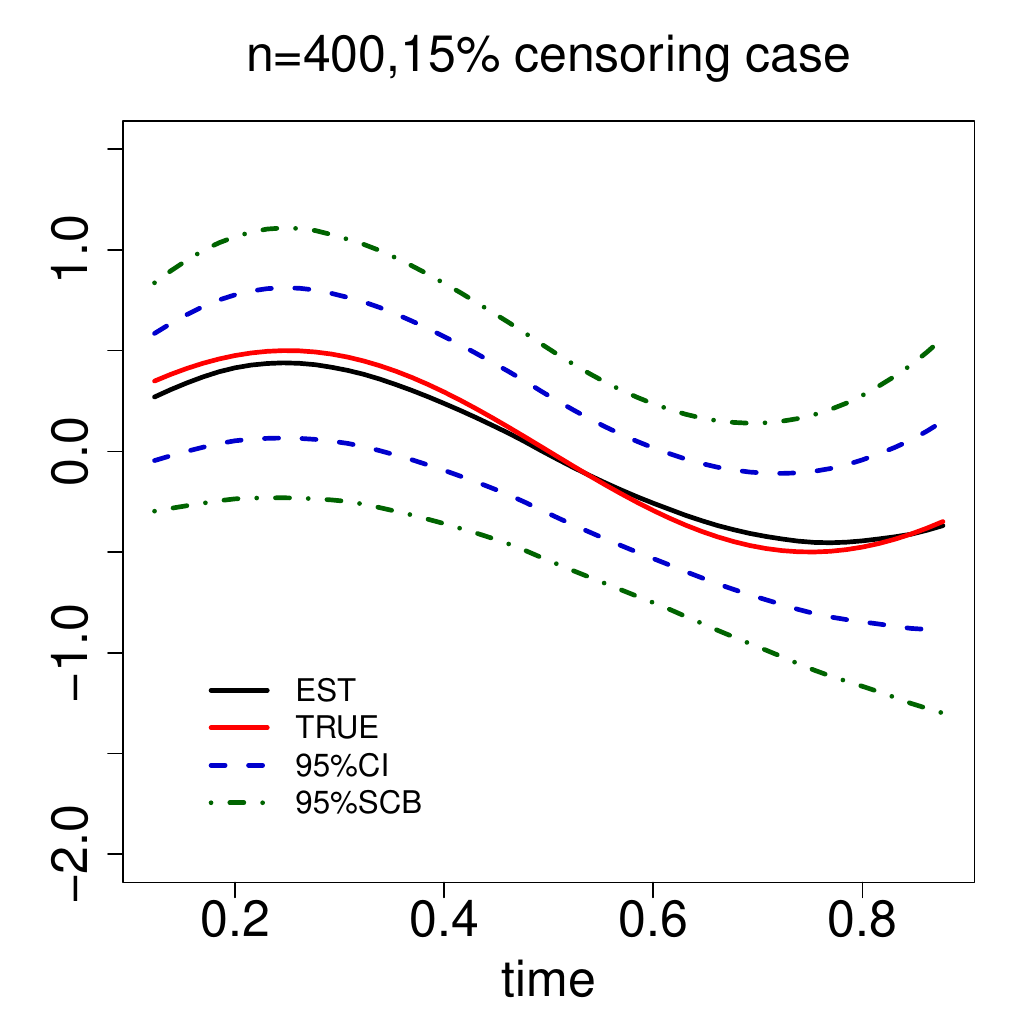}
\includegraphics[width=0.49\linewidth]{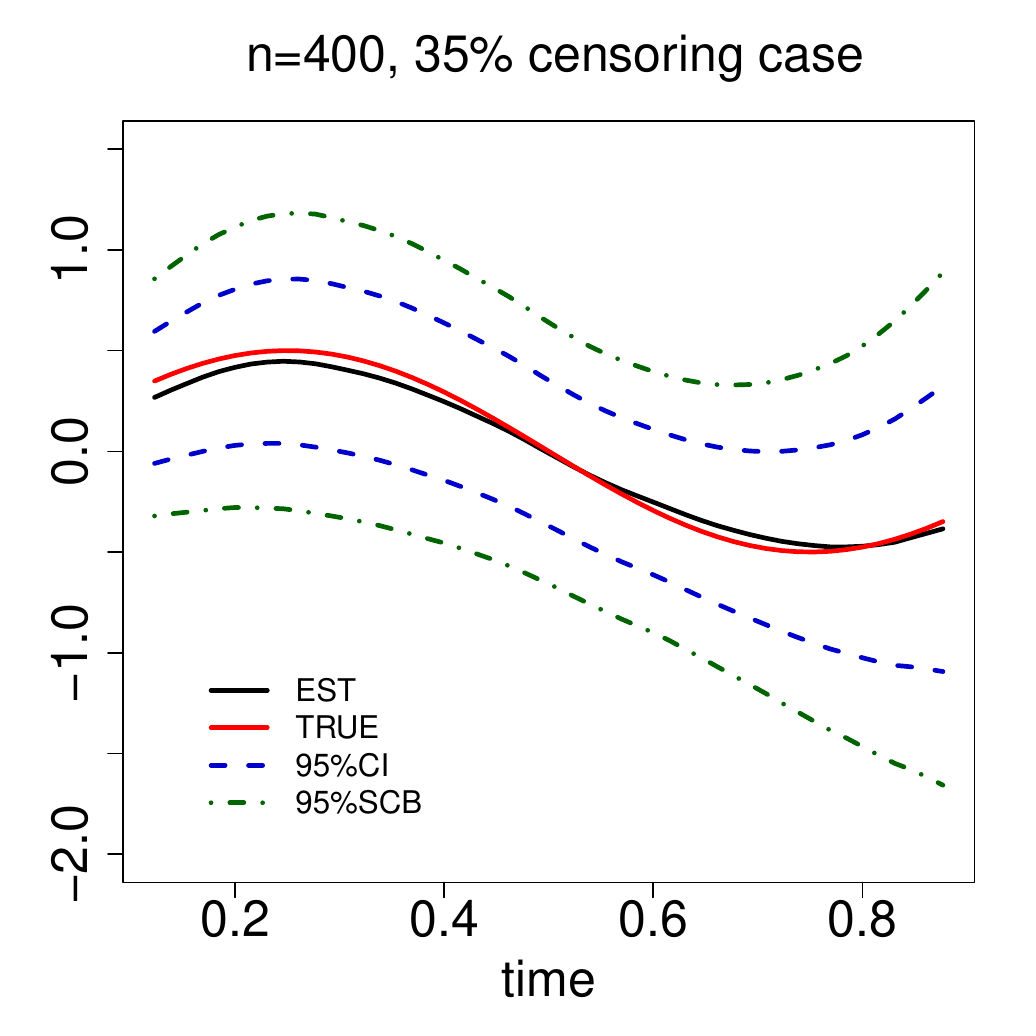}\\
\includegraphics[width=0.49\linewidth]{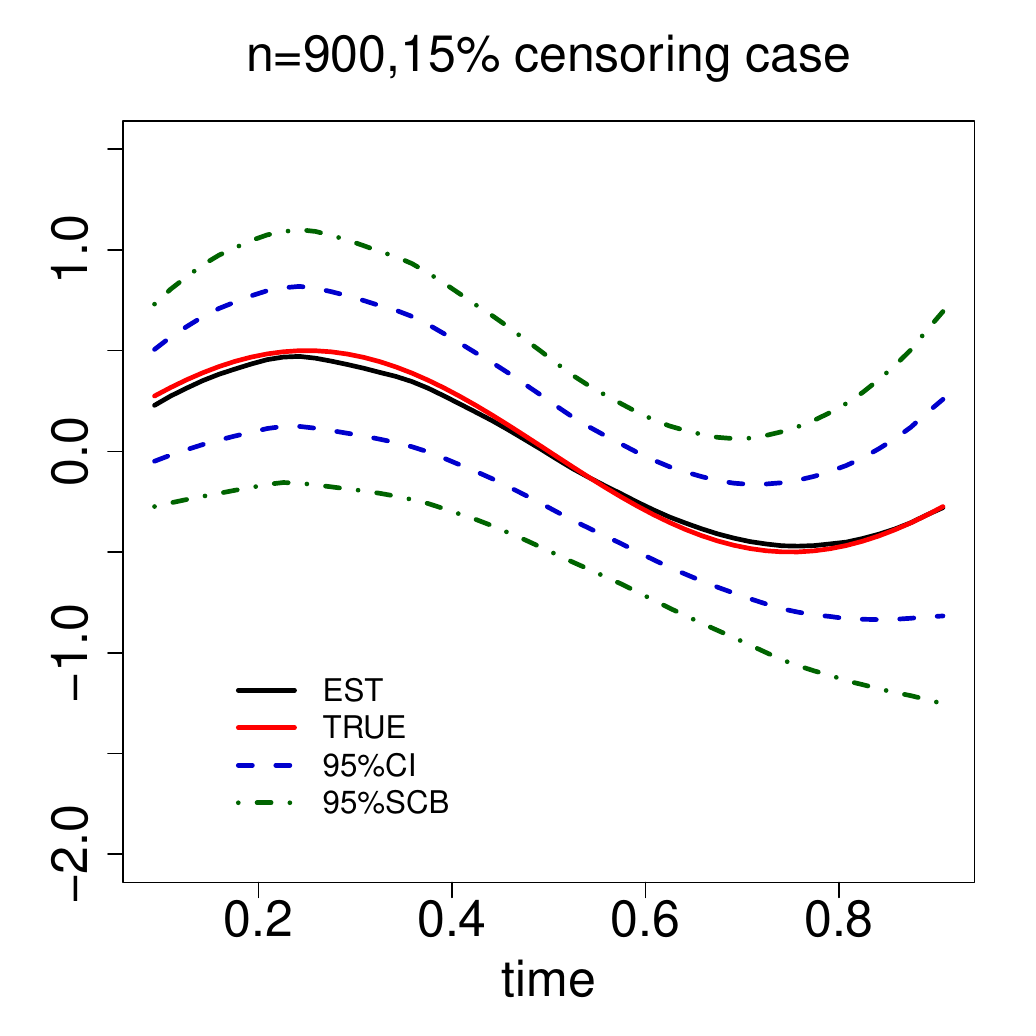}
\includegraphics[width=0.49\linewidth]{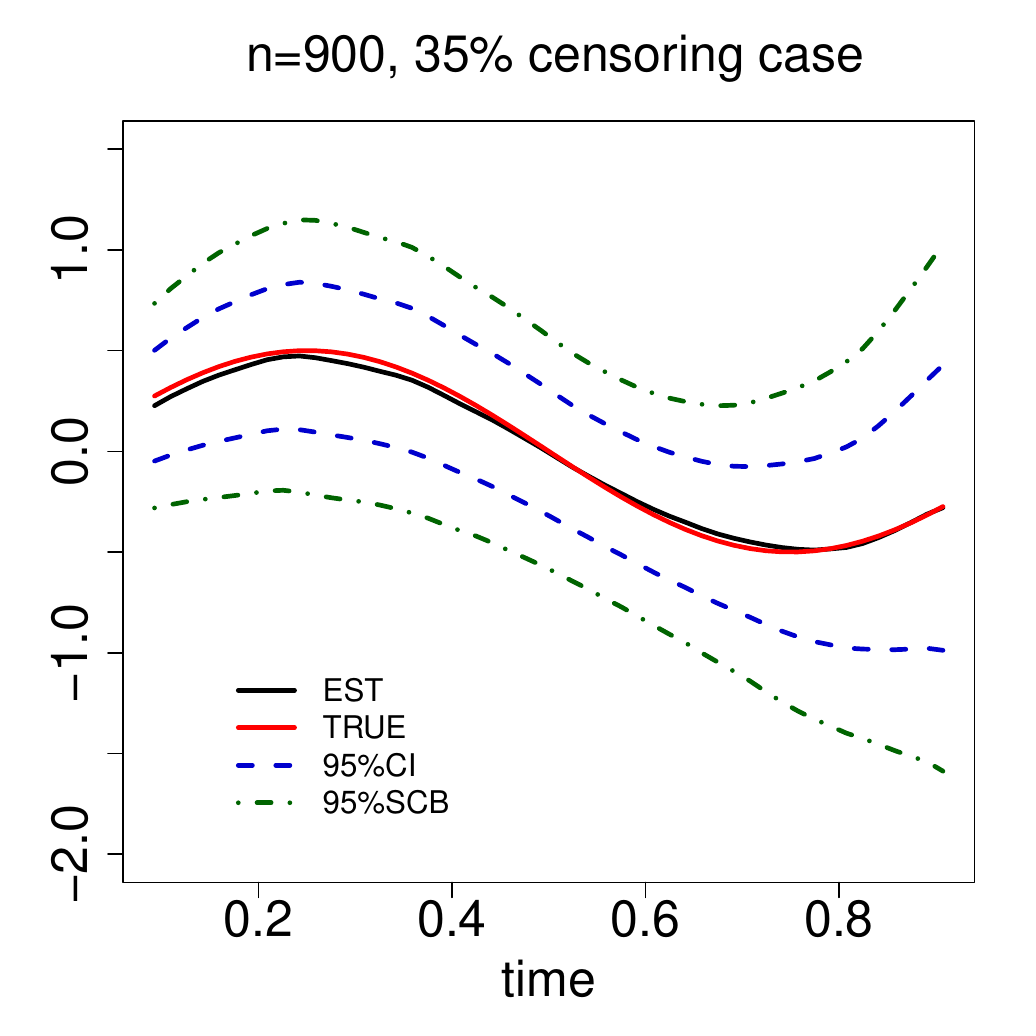}
\caption{Estimates with ``auto'' bandwidth selection approach.}
\label{fig:scb}
\end{center}
\end{figure}

 Figure \ref{fig:scb} shows the estimate $\hat{\beta}(s)$ 
 {using ``auto'' bandwidth selection rule with the sample sizes $400$ and $900$.} The left panel corresponds to $15\%$ censoring, and the right panel corresponds to $35\%$ censoring. 
 The green dashed curve is the $95\%$ SCB, and the blue dashed curve is the $95\%$ point-wise confidence interval.  
 {Figure \ref{fig:scb} shows that the SCBs become narrower and more accurate as the sample size increases.}
SCBs are wider than pointwise intervals, as they have uniform coverage throughout the time domain. 
 Table \ref{table:scb} summarizes the probability of uniform coverage of the SCBs and point-wise confidence intervals based on the simulations of $1000$ datasets. We observe that SCBs achieve coverage probabilities near the nominal level under different censoring rates. The performance improves with a larger sample size. On the other hand, the pointwise confidence interval is not valid for simultaneous inference, due to much lower uniform coverage probabilities.

\begin{table}[htbp]
\caption{Uniform coverage probability based on different methods}
{\label{table:scb}}
\begin{center}
\begin{tabular}{llcccccc}
\toprule
&  &   & \multicolumn{2}{c}{$15\%$\ Censoring\ rate\ \ } &  & \multicolumn{2}{c}{ $35\%$ Censoring\ rate\ \ }\\
\cline{4-5}   \cline{7-8}
$n$   &  Bandwidth &  & SCB & CI &  & SCB & CI \\
\midrule
400 & $h_1=n^{-0.35}, h_2=n^{-0.35}$  &  &  91.3 & 34.8  &  & 90.4  & 34.3 \\
    & $h_1=n^{-0.35}, h_2=n^{-0.45}$  &  &  92.0 & 32.5 &  & 90.5 & 29.3\\
     &  auto &  &  92.4 & 31.9 &  & 90.3 & 27.2\\
 900 & $h_1=n^{-0.35}, h_2=n^{-0.35}$ &  &  93.1 &  28.0  &  &  93.4 &  26.0\\
    & $h_1=n^{-0.35}, h_2=n^{-0.45}$  &  &  92.9 &  23.2 &  & 93.0 & 21.0\\
    & auto  &  &  93.8 & 17.6 &  & 91.9 & 16.6\\
\bottomrule
\end{tabular}
\end{center}
\footnotesize{Note: ``SCB"  refers SCB and ``CI" means point-wise confidence interval. } 
\end{table}

\section{Data Analysis} 
{We apply the proposed method to a dataset from the Alzheimer's Disease Neuroimaging Initiative (ADNI) study to demonstrate its practical application. ADNI is a large, ongoing 
study initiated in 2004. It supports the investigation and development of treatments that slow or stop the progresison of Alzheimer's Disease (AD), %
a progressive neurodegenerative disorder that affects memory and cognitive function.
We are interested in identifying possible risk factors and their dynamic effects for AD. Specifically, we look at APOE4 gene 
(coded as 0 for non-carriers, 1 for carriers of one allele, and 2 for carriers of two alleles), gender (coded as 1 for males and 0 for females), and the volumes of the hippocampus and entorhinal cortex (measured in cubic millimeters) on AD. While APOE4 and gender are time-independent, 
the volumes of the hippocampus and entorhinal cortex are measured longitudinally. } 

The dataset consists of 2430 subjects. 
After eliminating missing data, 2088 participants are used for analysis under missing at random assumption \citep{little2019}, 
among whom 736 ($35.2\%$) experienced AD. The admission time of participants is set as time origin and 
the maximum follow-up time $\tau=6280$ days. For simplicity, we set $h_1=h_2=h$. Then, we adopt the proposed automatic bandwidth selection method to 
select the bandwidth between $9(Q_3-Q_1)n^{-1/2}\approx 454$ days and $9(Q_3-Q_1)n^{-1/6}\approx 5798$ days, where  $Q_3$ is the 3rd quartile and $Q_1$ is the 1st quartile of the longitudinal measurement times, and $n$ is the total number of participants. Our theory suggests valid bandwidth should range from $O(n^{-1/2})$ to $o(n^{-1/6}).$ 
In practice, we need to decide the constants, which we use $9(Q_3-Q_1)$ here, producing a wide range of time for bandwidth selection.
We fit the data with the following varying-coefficient multiplicative hazards model:   
\begin{align*}
& \lambda\left\{t\mid  \text{APOE4},  \text{Gender}, \text{ Hippocampus} (r), \text{ Entorhinal} (r),r\leq t\right\}
=\lambda_0(t)\exp\left\{ \beta_1(t) \text{APOE4} \right.\\
&\left. + \beta_2(t) \text{Gender}+\beta_3(t)\text{Hippocampus}(t)+
\beta_4(t)\text{Entorhinal}(t)\right\}.
\end{align*}
For a fixed time point $s \in [h,\tau-h],$ solving (\ref{coxestequation}), 
we obtain $\hat{\beta}(s).$ Its variance is estimated by the sandwich formula (\ref{sandwich}). 
A $95\%$ point-wise confidence interval can be constructed based on normal approximation. 
For the construction of a simultaneous confidence band, we use 
$\beta_1(\cdot)$ to illustrate. 
We generate
$M=5000$ realizations of $\{{\xi}_i \stackrel{i.i.d.}{\sim} \mbox{Exp}(1)-1, i=1,\ldots,2088\},$ where $\mbox{Exp}(1)$ represents the exponential distribution with mean 1, and plug them into (\ref{perturbation-cox}). 
After obtaining $\tilde{U}_n\{\hat{\beta}(s)\}$ and $I\{ \hat{\beta}(s)\},$ we 
use the inverse of the estimated standard error of $\hat{\beta}_1(s)$ as the weight $\hat{w}(s)$ in the calculation of 
$\{\tilde{S}^{(1)}_{SCB},\ldots,\tilde{S}^{(5000)}_{SCB}\}$ in (\ref{tilderscb}). 
 The SCB of $\beta_1(\cdot)$ is constructed using (\ref{equ:scb}).
 
\begin{figure}[htb!]
\begin{center}
\includegraphics[width=0.49\linewidth]{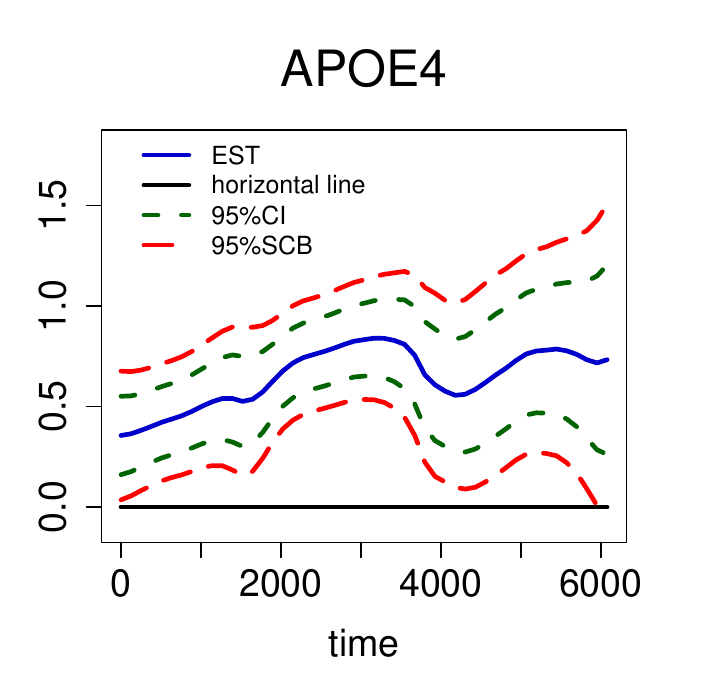}
\includegraphics[width=0.49\linewidth]{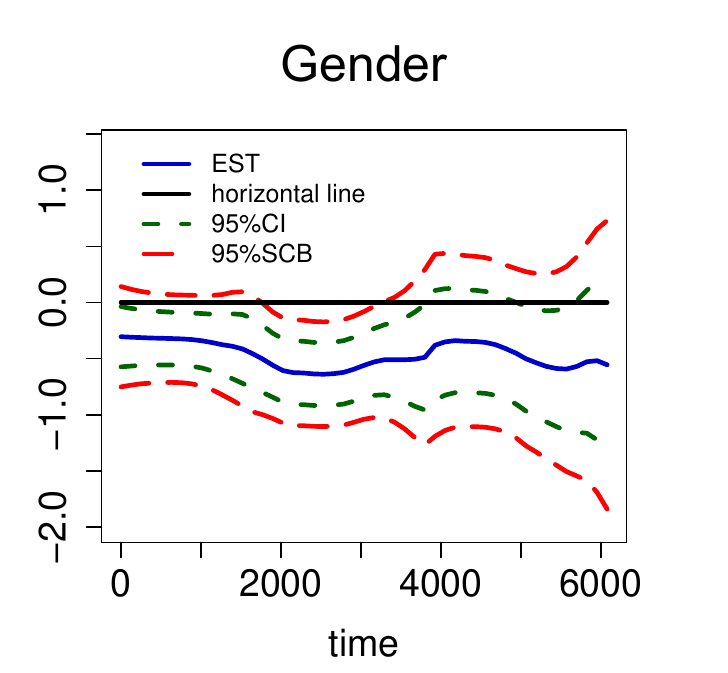}\\
\includegraphics[width=0.49\linewidth]{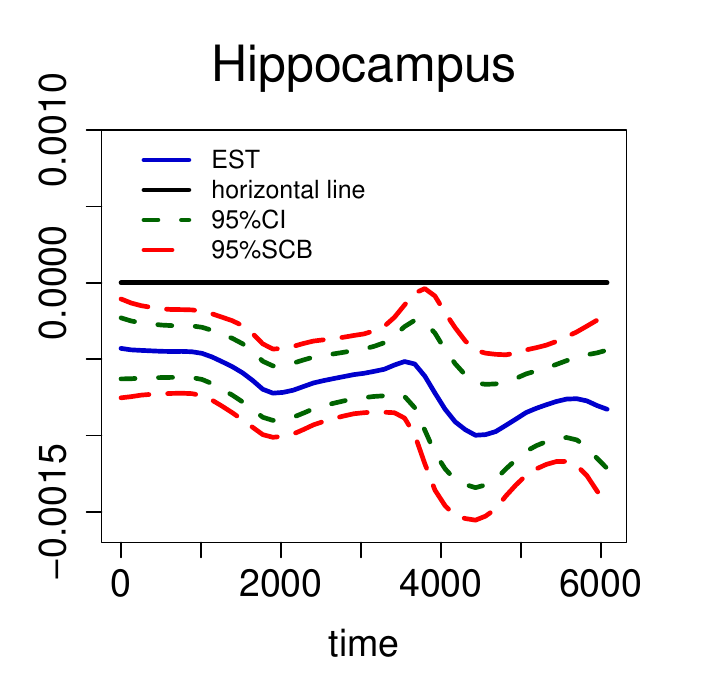}
\includegraphics[width=0.49\linewidth]{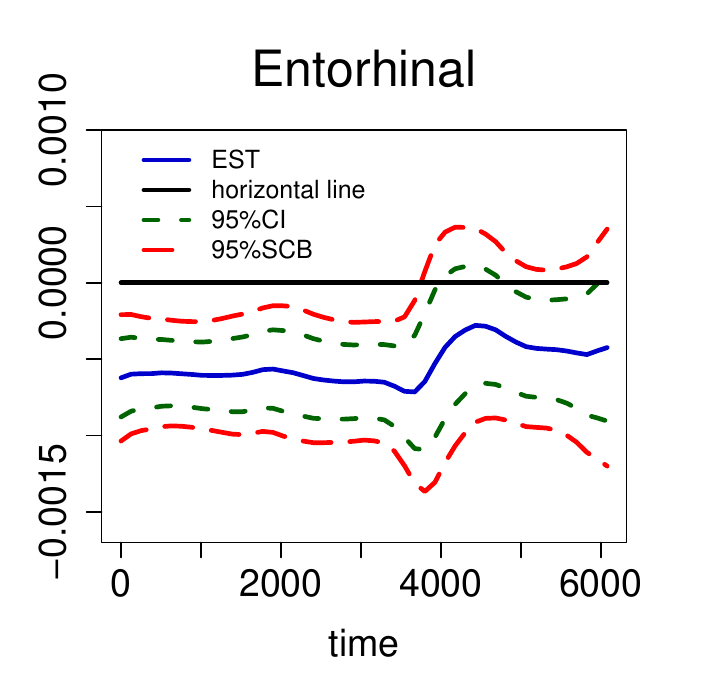}
\caption{Estimates with auto bandwidth selection approach.}
\label{r2:fig}
\end{center}
\end{figure}

We summarize the estimated coefficient function, $95\%$ point-wise confidence interval, and $95\%$ simultaneous confidence band in Figure \ref{r2:fig}. We also plot a horizontal line to represent the $0$ effect. Figure \ref{r2:fig} shows that carrying the APOE4 gene significantly increases the hazard of developing AD, with the hazard progressively rising over time. 
This is consistent with the literature that APOE4 allele is a genetic risk factor for AD, with carriers of this allele having a higher likelihood of developing the disease \citep{Yamazaki2019, Ding2024}. Our analysis reveals that the detrimental effect of carrying APOE4 increases over time. 
The APOE4 gene contributes to AD pathogenesis by impairing amyloid-beta clearance, increasing its aggregation, and promoting neuroinflammation. Over time, the cumulative effects lead to a higher hazard of developing AD. 
Furthermore, Figure \ref{r2:fig} suggests that women are more susceptible to AD than men, the effect is significant and 
stable over time. As highlighted by \cite{Pike2017} and \cite{Cui2023}, women exhibit a higher susceptibility to AD, attributed to both biological and hormonal factors. For instance, the loss of estrogen after menopause may exacerbate neurodegeneration and amyloid-beta pathology. 
Additionally, from Figure \ref{r2:fig}, we observe that a reduction  in hippocampal volume is associated with a higher risk of AD, with the 
effect increasing over time.  The hippocampus, a critical brain region involved in memory formation, is often one of the first areas to be affected by AD, leading to memory loss.  A decrease in hippocampal volume reflects neuronal loss and synaptic degeneration, hallmark features of AD \citep{Huijbers2020}. This structural decline progressively disrupts cognitive function, explaining the increased risk of AD over time as hippocampal volume decreases.
Finally, Figure \ref{r2:fig} indicates that a reduction in entorhinal cortex volume is linked to an elevated risk of AD, consistent with the findings of \cite{Tran2022}. The entorhinal cortex is involved in memory and spatial navigation and shows early signs of degeneration in AD, contributing to cognitive decline.  
We observe that the effect is more pronounced between 0--4000 days but tends to diminish between 4000--6000 days. The pronounced risk associated with its volume reduction in the earlier time frame (0--4000 days) may reflect its early involvement in the disease process, where pathological changes in this region trigger broader neurodegenerative cascades. The subsequent diminishment of risk (4000--6000 days) could suggest that by this stage, the disease has progressed to affect other brain regions, thereby distributing the risk factors more diffusely.

\section{Concluding Remarks}\label{conclude.sec}
In this paper, we propose to estimate the time-varying effect of longitudinally collected covariates for the multiplicative hazards model. This allows us to examine the dynamic relationship between time-dependent covariates and time-to-event outcomes. For any fixed time point, we establish the asymptotic normality of the proposed estimator. To quantify the uncertainty of the non-parametric coefficient function, we further develop a simultaneous confidence band through multiplier bootstrap. Simulation studies demonstrate the favorable performance of the proposed method, and an analysis of a dataset from Alzheimer's Disease Neuroimaging Initiative study reveals the dynamic relationship between APOE4 gene, gender, hippocampus volume and entorhinal cortex volume and the onset of AD.

We assume that the observational time of the covariate process is external. Our approach can be extended to the case of informative observational times, which may depend on the past covariate as in \cite{cao2016}. 
{We assume that the observation times of each individual occur at random and we aggregate information across different individuals. Asymptotically, there would be enough data at any time point $s$ as we increase the sample size. In finite samples, it is possible that for a particular point of interest $s,$ we do not have nearby observations to perform kernel smoothing.}
Our approach can be used to analyze the censored outcome with other models, such as the additive hazards model or the transformed hazards model \citep{sun2022, sun2023}. We leave these for future work.
In practice, some covariates may have a time-independent coefficient, and some covariates may have a time-dependent coefficient, which warrants more research.

\section*{Acknowledgement}
We would like to express our gratitude to the Editor, the Associate Editor, and two reviewers for their invaluable insights that significantly improved the quality of this manuscript. Cao's  research is partially supported by NSF DMS-2311249.

\bibliographystyle{apalike}
\bibliography{tdcox}

  \newpage
\appendix
\section*{Supplementary material}
This supplementary material provides proofs of Theorem \ref{th:coxnormal} and Corollary \ref{cor1} in section~\ref{supp:proof}. 
Section~\ref{supp:3} provides theoretical support for the resampling strategy used. Section~\ref{sec:c} presents additional simulation results.

\appendix
\section{Proofs of main results}
\label{supp:proof}
This section provides details on the proofs of Theorem \ref{th:coxnormal} and Corollary \ref{cor1}. Our main tool is the empirical processes. The proofs use two lemmas, which are stated and proved as follows.

\begin{lemma}\label{le:part2}
Under the conditions of Theorem \ref{th:coxnormal}, we have
\begin{align*}
&(nh_1h_2)^{1/2}\E\left(\int_0^{\tau}\int  K_{h_1,h_2}(t_1-t,t_2-t)\left[Z(t_2)-\bar{Z}\{\beta(t),t_1\}\right]d N^*_i(t_2) d N_i(t_1)\right)\\
&=(nh_1h_2)^{1/2}B\{\beta_0(t),t\}\big\{\beta(t)-\beta_0(t)\big\}+(nh_1h_2)^{1/2}O_p(h_1^2+h_1h_2+h_2^2),
\end{align*}
where
\begin{align*}
B\{\beta_0(t),t\}&=-\left[{s}^{(2)}\{\beta_0(t),t\}-\frac{{s}^{(1)}\{\beta_0(t),t\}^{\otimes 2}}{{s}^{(0)}\{\beta_0(t),t\}}\right]\lambda_0(t).
\end{align*}
\end{lemma}

\begin{lemma}\label{le:var}
Under the conditions of Theorem \ref{th:coxnormal},
\begin{align*}
&{\var}\left(\iint(h_1h_2)^{1/2}K_{h_1,h_2}(t-s,r-s)\left[Z(r)-\frac{{S}^{(1)}\{\beta(s),t\}}{{S}^{(0)}\{\beta(s),t\}}\right]d N^*(r)d N(t) \right)& \\
&=\iint K(z_1,z_2)^2 d z_2 d z_1\left\{{s}^{(2)}\{\beta_0(s),s\}-\frac{{s}^{(1)}\{\beta(s),s\}^{\otimes2}}{{s}^{(0)}\{\beta(s),s\}}\right\}\lambda_0(s).
\end{align*}
\end{lemma}
\subsection{Proof of Lemma \ref{le:part2}}

\begin{proof}
Recall that $E\{d N^*(r)\}=\lambda^*(r),$  $\lambda\{t\mid Z(r),r\leq t\}=\lambda_0(t)e^{\beta_0(t)^TZ(t)},$ $$S^{(l)}\{\beta(s),t\}=\frac{1}{n}\sum_{j=1}^n \sum_{k=1}^{M_j} K_{h_1,h_2}(t-s,R_{jk}-s)Y_j(t)Z_j(R_{jk})^{\otimes l}\exp\{\beta(s)^T Z_j(R_{jk})\}$$ and $$ s^{(l)}\{\beta(t),t\}=\E\left[ Y(t)Z(t)^{\otimes l}\exp\left\{\beta(t)^T Z(t)\right\}\right]\lambda^*(t),$$ 
where $l=0,1,2.$

Under Condition \ref{con:count},   and using change of variables, we have
\begin{align*}
&D(s) \equiv (nh_1h_2)^{1/2} \E\left(\int_0^{\tau}\int K_{h_1,h_2}(t-s,r-s)\left[Z(r)-\frac{{S}^{(1)}\{\beta(s),t\}}{{S}^{(0)}\{\beta(s),t\}}\right]d N^*(r) d N(t)\right)\\
&=(nh_1h_2)^{1/2} \Bigg(\iint K(t_1,t_2)E\left[Z(s+t_2h_2)Y(s+t_1h_1)e^{\beta_0(s+t_1h_1)^T Z(s+t_1h_1)} \right]   \\
&-\iint K(t_1,t_2)E\left[ \frac{{S}^{(1)}\{\beta(s),s+t_1h_1\}}{{S}^{(0)}\{\beta(s),s+t_1h_1\}} Y(s+t_1h_1)e^{\beta_0(s+t_1h_1)^T Z(s+t_1h_1)}\right]\Bigg)\\
&\times \lambda(s+t_1h_1,s+t_2h_2)dt_1dt_2, 
\end{align*}
where $\lambda(t_1,t_2)=\lambda_0(t_1)\lambda^*(t_2).$

Since $\int K(x) dx=1, \int xK(x)dx=0,$
after change of variables, we obtain
\begin{align*}
 D(t)& =(nh_1h_2)^{1/2}\E\left[Z(t)Y(t)e^{\beta_0(t)^TZ(t)} - \frac{S^{(1)}\{\beta(t),t\}}{S^{(0)}\{\beta(t),t\}}Y(t)e^{\beta_0(t)^TZ(t)}\right] \lambda(t,t)\\
&+(nh_1h_2)^{1/2}O_p(h_1^2+h_1h_2+h_2^2).\\
&=(nh_1h_2)^{1/2}\E\left[Z(t)Y(t)e^{\beta_0(t)^TZ(t)} - \frac{s^{(1)}\{\beta(t),t\}}{s^{(0)}\{\beta(t),t\}}Y(t)e^{\beta_0(t)^TZ(t)}+o_p(1)\right] \lambda(t,t)\\
&+(nh_1h_2)^{1/2}O_p(h_1^2+h_1h_2+h_2^2).
\end{align*}
For any fixed time point $t\in[h,\tau-h]$, we have 
\begin{align*}
D(t) &=-(nh_1h_2)^{1/2} \left[{s}^{(2)}\{\beta_0(t),t\}-\frac{{s}^{(1)}\{\beta_0(t),t\}^{\otimes 2}}{{s}^{(0)}\{\beta_0(t),t\}}\right]\lambda_0(t)\{\beta(t)-\beta_0(t)\}  \\
&+(nh_1h_2)^{1/2}O_p(h_1^2+h_1h_2+h_2^2)+o_p\left\{(nh_1h_2)^{1/2}\left|\beta(t)-\beta_0(t)\right|\right\}\\
&= (nh_1h_2)^{1/2}B\{\beta_0(t),t\}\{\beta(t)-\beta_0(t)\} +(nh_1h_2)^{1/2}O_p(h_1^2+h_1h_2+h_2^2)\\
& +o_p\{(nh_1h_2)^{1/2}\left|\beta(t)-\beta_0(t)\right| \},
\end{align*}
where
\begin{align*}
B\{\beta_0(t),t\}&=-\left[{s}^{(2)}\{\beta_0(t),t\}-\frac{{s}^{(1)}\{\beta_0(t),t\}^{\otimes 2}}{{s}^{(0)}\{\beta_0(t),t\}}\right]\lambda_0(t)\\
&=-E\left(Y(t)\exp\{\beta_0(t)^TZ(t)\}\left[Z(t)-\frac{s^{(1)}\{\beta_0(t),t\}}{s^{(0)}\{\beta_0(t),t\}}\right]\right)\lambda(t,t).
\end{align*}
 Under Condition \ref{con:var}, $B\left\{\beta_0(t),t\right\}$ is non-singular. It is a non-positive definite matrix.
\end{proof}

\subsection{ Proof of Lemma \ref{le:var}}
 
\begin{proof}
The variance can be calculated as follows:
\begin{align*}
&\Sigma=\var\left(\iint(h_1h_2)^{1/2}K_{h_1,h_2}(t-s,r-s)\left[Z(r)-\frac{
{S}^{(1)}\{\beta(s),t\}}{
{S}^{(0)}\{\beta(s),t\}}\right]d N^*(r)d N(t) \right) \\
&=  h_1h_2 \E\Bigg(\iiiint K_{h_1,h_2}(t_1-s,r_1-s)K_{h_1,h_2}(t_2-s,r_2-s)\left[Z(r_1)-\frac{
{S}^{(1)}\{\beta(s),t_1\}}{
{S}^{(0)}\{\beta(s),t_1\}} \right]\Bigg.\\
&\Bigg.\left[Z(r_2)-\frac{
{S}^{(1)}\{\beta(s),t_2\}}{
{S}^{(0)}\{\beta(s),t_2\}} \right]d N^*(r_1)d N(t_1)d N^*(r_2)d N(t_2)\Bigg)\\
&-\left\{\E\left(\iint(h_1h_2)^{1/2}K_{h_1,h_2}(t-s,r-s)\left[Z(r)-\frac{
{S}^{(1)}\{\beta(s),t\}}{
{S}^{(0)}\{\beta(s),t\}} \right]d N^*(r)d N(t) \right)\right\}^2\\
&=I_1-I_2.
\end{align*}
 For $I_2$, under \ref{con:continuous}, \ref{con:kernel} and \ref{con:covariate}, we have
\begin{align*}
&I_2=\left\{\E\left(\iint(h_1h_2)^{1/2}K_{h_1,h_2}(t-s,r-s)\left[Z(r)-\frac{
{S}^{(1)}\{\beta(s),t\}}{
{S}^{(0)}\{\beta(s),t\}} \right]d N^*(r)d N(t) \right)\right\}^2\\
&=(h_1h_2)\left(\iint K(z_1,z_2)E\left[Z(s+h_2z_2)Y(s+h_1z_1)e^{\beta_0(s+h_1z_1)^TZ(s+h_1z_1)} \right]\right.\\
&  \times\lambda(s+h_1z_1,s+h_2z_2)dz_1dz_2  \\
&-\iint K(z_1,z_2)\E\left[\frac{{S}^{(1)}\{\beta(s),s+h_2z_2\}}{{S}^{(0)}\{\beta(s),s+h_2z_2\}}Y(s+h_1z_1)e^{\beta_0^T(s+h_1z_1)Z(s+h_1z_1)}\right]\\
 &  \times\lambda(s+h_1z_1,s+h_2z_2)dz_1dz_2 \bigg)^2\\
&=(h_1h_2)\Big(\left[{s}^{(1)}\{\beta_0(s),s\}-{s}^{(1)}\{\beta_0(s),s\}+o_p(1)\right]\lambda_0(t)+O_p(h_1^2+h_1h_2+h_2^2)\Big)^2=o_p(h_1h_2).
\end{align*}

Then we decompose $I_1$ into four parts.
\begin{align*}
&I_1=\E\left(h_1h_2 \iiiint_{t_1\ne t_2\atop r_1\ne r_2}  K_{h_1,h_2}(t_1-s,r_1-s)K_{h_1,h_2}(t_2-s,r_2-s)\right.\\
&\left.\left[Z(r_1)-\frac{{S}^{(1)}\{\beta(s),t_1\}}{{S}^{(0)}\{\beta(s),t_1\}}\right]
\left[Z(r_2)-\frac{{S}^{(1)}\{\beta(s),t_2\}}{{S}^{(0)}\{\beta(s),t_2\}}\right]d N^*(r_1)dN(t_1)d N^*(r_1)d N(t_2)\right)\\
&+\E\left(h_1h_2 \iiint_{t_1\ne t_2}  K_{h_1,h_2}(t_1-s,r-s)K_{h_1,h_2}(t_2-s,r-s)\left[Z(r)-\frac{{S}^{(1)}\{\beta(s),t_1\}}{{S}^{(0)}\{\beta(s),t_1\}}\right] \right.\\
&\left. \times \left[Z(r)-\frac{{S}^{(1)}\{\beta(s),t_2\}}{{S}^{(0)}\{\beta(s),t_2\}}\right]d N(t_1)d N^*(r)d N(t_2)\right)\\
&+\E\left(h_1h_2 \iiint_{ r_1\ne r_2}  K_{h_1,h_2}(t-s,r_1-s)K_{h_1,h_2}(t-s,r_2-s)\left[Z(r_1)-\frac{{S}^{(1)}\{\beta(s),t\}}{{S}^{(0)}\{\beta(s),t\}}\right]\right.\\
&\left. \left[Z(r_2)-\frac{{S}^{(1)}\{\beta(s),t\}}{{S}^{(0)}\{\beta(s),t\}}\right] d N^*(r_1)d N^*(r_2)d N(t)\right)\\
&+\E\left(h_1h_2 \iint K_{h_1,h_2}(t-s,r-s)^2\left[Z(r)-\frac{{S}^{(1)}\{\beta(s),t\}}{{S}^{(0)}\{\beta(s),t\}}\right] ^{\otimes2} d N^*(r)d N(t)\right)\\
&:=I+II+III+IV.
\end{align*}
Using change of variables and \ref{con:count}, \ref{con:continuous} and \ref{con:kernel}, it can be shown that $I=O(h_1h_2)$, $II=O(h_1)$,  and $III=O(h_2)$. Hence, $IV$ is the main term and
\begin{align*}
IV  &=h_1h_2\E\left( \iint (h_1h_2)^{-2}K(z_1,z_2)^2\left[Z(s+z_2h_2)-\frac{{S}^{(1)}\{\beta(s),s+z_1h_1\}}{{S}^{(0)}\{\beta(s),s+z_1h_1\}}\right] ^{\otimes2}\right. \\
    & \times \lambda^*(s+z_2h_2)h_2Y(s+z_1h_1)\lambda_0(s+z_1h_1)e^{\beta_0(s+z_1h_1)^TZ(s+z_1h_1)}h_1d z_2d z_1\bigg)\\
    &=\iint K(z_1,z_2)^2\E\left(\left[Z(s+z_2h_2)-\frac{{S}^{(1)}\{\beta(s),s+z_1h_1\}}{{S}^{(0)}\{\beta(s),s+z_1h_1\}}\right] ^{\otimes2}\right. \\
    &\times \lambda^*(s+z_2h_2)Y(s+z_1h_1)e^{\beta_0(s+z_1h_1)^TZ(s+z_1h_1)}\lambda_0(s+z_1h_1)\bigg)d z_2d z_1\\
    &=\iint K(z_1,z_2)^2\left[ {s}^{(2)}\{\beta_0(s),s\}-\frac{{s}^{(1)}\{\beta_0(s),s\}^{\otimes2}}{{s}^{(0)}\{\beta_0(s),s\}}\right]\lambda_0(s)d z_2d z_1+O(h_1^2+h_1h_2+h_2^2).
\end{align*}
Therefore, when $n\rightarrow \infty,$ we have
\begin{align*}
\Sigma\left(\beta_0\left(s\right),s \right)=\iint K(z_1,z_2)^2 d z_2 d z_1\left[{s}^{(2)}\{\beta_0(s),s\}-\frac{{s}^{(1)}\{\beta_0(s),s\}^{\otimes2}}{{s}^{(0)}\{\beta_0(s),s\}}\right]\lambda_0(s).
\end{align*}
\end{proof}

\subsection{ Proof of Theorem \ref{th:coxnormal}}
\begin{proof}
Before the proof, additional notations are needed. Denote
$$
u\{\beta(t),\beta_0(t),t\}=\left[s^{(1)}\{\beta_0(t),t\}-\frac{s^{(1)}\{\beta(t),t\}}{s^{(0)}\{\beta(t),t\}}s^{(0)}\{\beta_0(t),t\}  \right]\lambda_0(t),
$$
and
$$
v\{\beta(t),\beta_0(t),t\}=-\left[ {{s}^{(2)}\{\beta(t),t\}} -\frac{{s}^{(1)}\{\beta(t),t\}^{\otimes 2}}{{s}^{(0)}\{\beta(t),t\}}\right]\frac{{s}^{(0)}\{\beta_0(t),t\}}{{s}^{(0)}\{\beta(t),t\}}\lambda_0(t).
$$
We can rewrite $v\{\beta(t),\beta_0(t),t\}$ as 
$$
-E\left(Y(t)\exp\{\beta(t)^TZ(t)\}\left[Z(t)-\frac{{s}^{(1)}\{\beta(t),t\}}{{s}^{(0)}\{\beta(t),t\}}\right]^{\otimes2}\frac{{s}^{(0)}\{\beta_0(t),t\}}{{s}^{(0)}\{\beta(t),t\}}\right)\lambda(t,t).
$$
For a fixed time point $t \in [h,\tau-h]$ and $\beta_0(t),$ to show the consistency of $\hat{\beta}(t),$ we first need to verify that  $u\{\beta(t),\beta_0(t),t\}=0$  when  $\beta(t)={\beta}_0(t).$ This follows by the definition of  $u\{\beta(t),\beta_0(t),t\}.$ Next,  from Condition \ref{con:var} and that $v\{\beta(t),\beta_0(t),t\}$ is a negative semi-definite matrix for any $\beta(t),$ it follows that ${\beta}_0(t)$ is the unique root to $u\{\beta(t),\beta_0(t),t\}=0$. Finally, we show that $U_n\{\beta(t)\}$ converges in probability to $u\{\beta(t),\beta_0(t),t\}$.

By \ref{con:count}, \ref{con:continuous}, \ref{con:kernel} and \ref{con:covariate}, 
we have 
\begin{align*}
&  E\left[U_n\{\beta(t)\}\right] \\
&=\frac{1}{n}\sum_{i=1}^nE\bigg(\int_0^{\tau}\int K_{h_1,h_2}(t_1-t,t_2-t)\left[Z_i(t_2)-\bar{Z}\{\beta(t),t_1\}\right]  dN^*_i(t_2) d N_i(t_1)\bigg)\\
&=\frac{1}{n}\sum_{i=1}^nE\bigg(\int_0^{\tau}\int K(z_1,z_2)\left[Z_i(t+z_2h_2)-\bar{Z}\{\beta(t),t+z_1h_1\}\right]  \lambda^*_i(t+z_2h_2)\\
&\times Y_i(t+z_1h_1) \lambda_0(t+z_1h_1)e^{\beta_0(t+z_1h_1)^TZ_i(t+z_1h_1)}dz_2 dz_1 \bigg)\\
&=\frac{1}{n}\sum_{i=1}^nE\bigg(\int_0^{\tau}\int K(z_1,z_2)\left[Z_i(t)-\bar{Z}\{\beta(t),t\}\right]  \lambda^*_i(t) Y_i(t) \lambda_0(t)e^{\beta_0(t)^TZ_i(t)}dz_2 dz_1\bigg) \\
&+O_p(h_1^2+h_1h_2+h_2^2)\\
&=\left[s^{(1)}\{\beta_0(t),t\}-\frac{s^{(1)}\{\beta(t),t\}}{s^{(0)}\{\beta(t),t\}}s^{(0)}\{\beta_0(t),t\}  \right]\lambda_0(t)+O_p(h_1^2+h_1h_2+h_2^2).
\end{align*}
By law of large numbers $U_n\{\beta(t)\} \to u\{\beta(t),\beta_0(t),t\}$ in probability, as $n \to \infty.$ Thus,  under  \ref{con:count},\ref{con:continuous}, \ref{con:kernel}, \ref{con:covariate} and convex function theory in \cite{andersen1982}, $\hat{\beta}(t) \to {\beta}_0(t)$ in probability.

Next we show the asymptotic normality of the proposed estimator.
We first establish the following relationship:
\begin{align}\label{proof:keyidea}
&\sup\limits_{\left|\beta(t)-\beta_0(t)\right|<M(nh_1h_2)^{-1/2}}\bigg|(nh_1h_2)^{1/2}{U}_n\left\{\beta(t)\right\}-(nh_1h_2)^{1/2}B\{\beta_0(t),t\}\{\beta(t)-\beta_0(t)\}\bigg.\notag\\
&-\left.(nh_1h_2)^{1/2}\left({U}_n\left\{\beta_0(t)\right\}-E\bigg[{U}_n\left\{\beta_0(t)\right\}\bigg]\right)\right|=(nh_1h_2)^{1/2}O_p(h_1^2+h_1h_2+h_2^2)\notag\\
&+o_p\{1+(nh_1h_2)^{1/2}\big|\beta(t)-\beta_0(t)\big|\}+o_p\{(nh_1h_2)^{1/2}(h_1^2+h_1h_2+h_2^2)\}.
\end{align}
To obtain (\ref{proof:keyidea}), first, using $\mathcal{P}_n$ and $\mathcal{P}$ to denote the empirical measure and true probability measure respectively, we have
\begin{align} \label{proof:Un}
&(nh_1h_2)^{1/2}{U}_n\{\beta(t)\}=(nh_1h_2)^{1/2}(\mathcal{P}_n-\mathcal{P})\left(\int_0^{\tau}\int K_{h_1,h_2}(t_1-t,t_2-t) \right.\notag\\
& \times \left[Z(t_2)-\bar{Z}\{\beta(t),t_1\}\right]d N^*_i(t_2) d N_i(t_1)\bigg)\notag\\
&+(nh_1h_2)^{1/2}E\bigg(\int_0^{\tau}\int K_{h_1,h_2}(t_1-t,t_2-t)\left[Z(t_2)-\bar{Z}\{\beta(t),t_1\}\right]  dN^*_i(t_2) d N_i(t_1)\bigg)\notag\\
&=I_1+I_2
\end{align}
 For the second term on the right-hand side of (\ref{proof:Un}), using Lemma \ref{le:part2}, we have
\begin{align}\label{proof:B}
I_2&=(nh_1h_2)^{1/2}B\left\{\beta_0(t),t\right\}\big\{\beta(t)-\beta_0(t)\big\}+ (nh_1h_2)^{1/2}O_p(h_1^2+h_1h_2+h_2^2)\notag\\
& +o_p\left\{(nh_1h_2)^{1/2}\left|\beta(t)-\beta_0(t)\right|\right\}.
\end{align}
From Lemma \ref{le:part2}, we know that $B\left\{\beta_0(t),t\right\}$ is a non-positive definite matrix. For the first term on the right-hand side of (\ref{proof:Un}), at fixed time point $t$, we consider the class of functions
$$
\left\{(h_1h_2)^{1/2}\int_0^{\tau}\int K_{h_1,h_2}(t_1-t,t_2-t)\left\{Z(t_2)-\bar{Z}\{\beta(t),t_1\}\right\}d N^*_i(t_2) d N_i(t_1):\left|\beta(t)-\beta_0(t)\right|<\epsilon\right\}
$$
for a given constant $\epsilon.$ Note that the functions in this class are Lipschitz continuous in $\beta(t)$ and the Lipschitz constant in uniformly bounded by
$$
(h_1h_2)^{1/2} M_1,
$$
which has finite second moment and $M_{1}$ is the upper bound of $$\left\|\E\left( \int_0^{\tau}\int K_{h_1,h_2}(t_1-t,t_2-t)\left[\frac{S^{(2)}\{\beta(t),t_1\}}{S^{(0)}\{\beta(t),t_1\}}-\frac{S^{(1)}\{\beta(t),t_1\}^{\otimes2}}{S^{(0)}\{\beta(t),t_1\}}\right]d N^*_i(t_2) d N_i(t_1)\right)\right\|_{\infty}. $$ Therefore,
this class is P-Donsker class by Jain-Marcus theorem \citep{vaart1996}.
As a result, we obtain that the first term on the right-hand side of $(\ref{proof:Un})$ for $\left|\beta(t)-\beta_{0}(t)\right|<$
$M\left(n h_1h_2\right)^{-1 / 2}$ is equal to
\begin{align}\label{proof:mvt}
&(nh_1h_2)^{1/2}(\mathcal{P}_n-\mathcal{P})\left(\int_0^{\tau}\int K_{h_1,h_2}(t_1-t,t_2-t)\left[Z(t_2)-\bar{Z}\{\beta_0(t),t_1\}\right]d N^*_i(t_2) d N_i(t_1)\right) \notag\\
&+o_{p}(1) \notag\\
=&\left(n h_1h_2\right)^{1 / 2}\big({U}_n\big\{\beta_{0}(t)\big\}-E\left[{U}_n\big\{\beta_{0}(t)\big\}\right]\big)+o_{p}(1).
\end{align}

Combining (\ref{proof:B}) and (\ref{proof:mvt}), we obtain (\ref{proof:keyidea}). Let $\beta(t)=\hat{\beta}(t)$ in (\ref{proof:mvt}). Consequently,
\begin{align*}\label{proof:end}
&(nh_1h_2)^{1/2}B\{\beta_0(t),t\}\{\hat{\beta}(t)-\beta_0(t)\}+(nh_1h_2)^{1/2}O_p(h_1^2+h_1h_2+h_2^2)\notag\\
& +o_p\left\{1+(nh_1h_2)^{1/2}\left|\beta(t)-\beta_0(t)\right|\right\}\notag\\
&=(nh_1h_2)^{1/2}\big({U}_n\big\{\beta_{0}(t)\big\}-E\left[{U}_n\big\{\beta_{0}(t)\big\}\right]\big).
\end{align*}
On the other hand, from Lemma \ref{le:var}, we have
\begin{align*}
{\Sigma}(\beta_0,s)=\iint K(z_1,z_2)^2\left\{{s}^{(2)}(\beta_0,s)-\frac{{s}^{(1)}(\beta_0,s)}{{s}^{(0)}(\beta_0,s)} \right\} \lambda_0(s)dz_2dz_1.
\end{align*}
To prove the asymptotic normality, next, we verify  Lyapunov condition holds.  Define
\begin{align*}
\psi_{i} =\left(n h_1h_2\right)^{1 / 2} n^{-1} \iint K_{h_1,h_2}(t-s,r-s)\left[Z(r)  -\frac{{S}^{(1)}\left\{\beta_{0}(s), t\right\}}{{S}^{(0)}\left\{\beta_{0}(s), t\right\}}\right]
d N^{*}(r)  d N(t).
\end{align*}
Similar to the calculation of $\Sigma\left(\beta_{0},s\right)$,
\begin{align*}
\sum_{i=1}^{n} \E \left(\left|\psi_{i}-\E \psi_{i}\right|^{3}\right)
=n O\{(n h_1h_2)^{3 / 2} n^{-3} (h_1h_2)^{-2}\}=O\left\{\left(n h_1h_2\right)^{-1 / 2}\right\}.
\end{align*}
Thus,
$$
\left(n h_1h_2\right)^{1 / 2}\left({U}_n\big\{\beta_{0}(t)\big\}-E\left[{U}_n\big\{\beta_{0}(t)\big\}\right]\right) \rightarrow N\big(0, {\Sigma}\left(\beta_{0},s\right) \big).
$$
Combing with (\ref{proof:keyidea}), we finish the proof.
\end{proof}

\subsection{Proof of Corollary \ref{cor1}}
\begin{proof}
In this part, we will prove the consistency of the variance estimate.
First, for a fixed time point $s\in [h,\tau-h],$ we have
\begin{align*}
&-\frac{\partial {U}_n \{\beta(s) \}}{\partial \beta(s) } =&\\
&n^{-1} \sum_{i=1}^{n} \int_{0}^{\tau} \left[\int_{0}^{\infty} K_{h_1,h_2}(u-s,r-s) d N_{i}^{*}(r)\right]\left[\frac{{S}^{(2)}\{\beta(s), u\}}{{S}^{(0)}\{\beta(s), u\}}-\frac{{S}^{(1)}\{\beta(s), u\}^{\otimes 2}}{{S}^{(0)}\{\beta(s), u\}^{\otimes 2}}\right] d N_{i}(u). 
\end{align*}
Using the similar argument to obtain (\ref{proof:mvt}), we can prove that
$$
\left\{\iint K_{h_1,h_2}(u-s,r-s) d N^{*}(r)\left[\frac{{S}^{(2)}\{\beta(s), u\}}{{S}^{(0)}\{\beta(s), u\}}-\frac{{S}^{(1)}\{\beta(s), u\}^{\otimes 2}}{{S}^{(0)}\{\beta(s), u\}^{\otimes 2}}\right] d N(u):\left|\beta(s)-\beta_{0}(s)\right|<\epsilon\right\}
$$
is a P-Glivenko-Cantelli class.

As a result, for a fixed time point $s\in [h,\tau-h],$
$$\sup _{\left|\beta(s)-\beta_{0}(s)\right|<\epsilon}\left| \frac{\partial {U}_n\{\beta(s)\}}{\partial \beta(s)}\mid_{\beta(s)=\hat{\beta}(s)}-\E\{\frac{\partial {U}_n\{\beta(s)\}}{\partial \beta(s)}\mid_{\beta(s)=\hat{\beta}(s)}\} \right| \rightarrow 0$$ in
probability.

From Theorem \ref{th:coxnormal}, $\hat{\beta}(s)$ is consistent for $\beta_{0}(s)$, by continuous mapping theorem, $\left.\frac{\partial {U}_n\{\beta(s)\}}{\partial \beta(s)}\right|_{\beta(s)=\hat{\beta}(s)}$ converges in probability to $B\{\beta_{0}(s),s\} $ for a fixed time point $s\in [h,\tau-h].$
Similarly, let $$\hat{{\Sigma}}({\beta},s)=n^{-2} \sum_{i=1}^{n}\left( \int_{0}^{\tau} \int_{0}^{\infty} K_{h_1,h_2}(u-s,r-s)\left[Z_{i}(r) 
-\bar{Z}\{\beta(s), u\} \right] d N_{i}^{*}(r) d N_{i}(u)\right)^{\otimes 2},$$
then $\sup _{\left|\beta(s)-\beta_{0}(s)\right|<\epsilon}|\hat{{\Sigma}}\big({\beta},s\big)-\E\{{\hat{\Sigma}}\big({\beta},s\big)\}| \rightarrow 0$ in probability.

On the
other hand,
$$
\E \{\hat{{\Sigma}}\big(\beta_0,s\big)\}=n^{-1} {\E} \left(\int_{0}^{\tau} \int_{0}^{\infty} K_{h_1,h_2}(u-s,r-s)\left[Z_{i}(r)-\bar{Z}\{\beta(s), u\}\right] d N_{i}^{*}(r) d N_{i}(u)\right)^{\otimes 2}.
$$
After change of variables, with \ref{con:continuous} and \ref{con:covariate},
$$
\E\left\{\hat{{\Sigma}}\big(\beta_0,s\big)\right\}=\frac{1}{n h_1h_2} \iint  K(z_1,z_2)^{2}   \left\{{s}^{(2)}\left\{\beta_{0}(s), s\right\}-\frac{{s}^{(1)}\left\{\beta_{0}(s), s\right\}^{\otimes 2}}{{s}^{(0)}\left\{\beta_{0}(s), s\right\}}\right\}\lambda_0(s)d z_1 d z_2.
$$
Therefore,
$$
\left(n h_1h_2\right) \hat{{\Sigma}}\big(\beta_0,s\big) \stackrel{p}{\rightarrow} {\Sigma}\big(\beta_{0},s\big) \quad \text { as } \quad n h_1h_2 \rightarrow \infty.
$$
The consistency of variance estimate follows.
\end{proof}

\section{The validity proof of the resampling strategy}
\label{supp:3}
\begin{proof}
The key to verifying the validity of the proposed resampling strategy is to prove that the perturbation-based estimating equation $\tilde{U}_n\left\{\hat{\beta}(s)\right\}$ is asymptotically equivalent to the original estimating equation $U_n\left\{\hat{\beta}(s)\right\}$ conditional on the data $D_n=\{(X_i,\delta_i, Z_i(R_{ik}), R_{ik}, k = 1,\ldots, M_i), i = 1, \ldots, n\},$ where
$$  
 U_n\left\{\beta(s)\right\}=n^{-1}\sum_{i=1}^n\sum_{k=1}^{M_i}\int_0^{\tau}K_{h_1,h_2}(t-s,R_{ik}-s)\left[Z_i(R_{ik})-\bar{Z}\{{\beta}(s),t\}\right]  d N_i(t),
$$
and
$$ 
\tilde{U}_n\left\{\beta(s)\right\}= n^{-1}\sum_{i=1}^n\sum_{k=1}^{M_i}\int_0^{\tau}K_{h_1,h_2}(t-s,R_{ik}-s)\left[Z_i(R_{ik})-\bar{Z}\{{\beta}(s),t\}\right]\xi_{i}  dN_i(t).
$$
For a fixed time point $s \in [h,\tau-h],$ we consider $\{\xi_i, i=1,\ldots,n\}$ as the only random component and $\{(X_i,\delta_i, Z_i(R_{ik}), R_{ik}, k = 1,\ldots, M_i), i = 1, ..., n\}$ as fixed in $\tilde{U}_n\left\{\beta(s)\right\}.$ Then the perturbation-based estimating equation $\tilde{U}_n\left\{\beta(s)\right\}$ can be regarded as a summation of $n$ independent random vectors. 

First, we calculate the conditional expectation of $\tilde{U}_n\left\{\hat{\beta}(s)\right\}:$
\begin{align*}
    & E\left[ n^{-1}\sum_{i=1}^n\sum_{k=1}^{M_i}\int_0^{\tau}K_{h_1,h_2}(t-s,R_{ik}-s)\left[Z_i(R_{ik})-\bar{Z}\{\hat{\beta}(s),t\}\right]\xi_{i}  dN_i(t) \Big| D_n\right] \\
    &=n^{-1}\sum_{i=1}^n\sum_{k=1}^{M_i}\int_0^{\tau}K_{h_1,h_2}(t-s,R_{ik}-s)\left[Z_i(R_{ik})-\bar{Z}\{\hat{\beta}(s),t\}\right]   dN_i(t)  E\left[\xi_{i}  \mid D_n\right]\\
    &=0.
\end{align*}
Second, the conditional variance of $\tilde{U}_n\left\{\hat{\beta}(s)\right\}$ is as follows.
\begin{align*}
    & n^{-2}\sum_{i=1}^n E \left[ \iiiint K_{h_1,h_2}(t_1-s,r_1-s)K_{h_1,h_2}(t_2-s,r_2-s)\left[Z_i(r_1)-\bar{Z}\{{\beta}(s),t_1\}\right] \right.\\
    &\times \left. \left[Z_i(r_2)-\bar{Z}\{{\beta}(s),t_2\}\right]\xi_{i}^2  dN_i^*(r_1)  dN_i(t_1)dN_i^*(r_2)  dN_i(t_2) \Big| D_n\right] \\
    &=n^{-2}\sum_{i=1}^n  \iiiint K_{h_1,h_2}(t_1-s,r_1-s)K_{h_1,h_2}(t_2-s,r_2-s)\left[Z_i(r_1)-\bar{Z}\{{\beta}(s),t_1\}\right]  \\
    &\times  \left[Z_i(r_2)-\bar{Z}\{{\beta}(s),t_2\}\right]dN_i^*(r_1)  dN_i(t_1)dN_i^*(r_2)  dN_i(t_2) E \left[ \xi_{i}^2   \mid D_n\right] \\
     &=n^{-2}\sum_{i=1}^n  \iiiint K_{h_1,h_2}(t_1-s,r_1-s)K_{h_1,h_2}(t_2-s,r_2-s)\left[Z_i(r_1)-\bar{Z}\{{\beta}(s),t_1\}\right]  \\
    &\times  \left[Z_i(r_2)-\bar{Z}\{{\beta}(s),t_2\}\right]dN_i^*(r_1)  dN_i(t_1)dN_i^*(r_2)  dN_i(t_2).
\end{align*}
Following Lemma \ref{le:var}, the conditional variance of $\sqrt{nh_1h_2}\tilde{U}_n\left\{\hat{\beta}(s)\right\}$ converges to the variance of  $\sqrt{nh_1h_2}{U}_n\left\{\hat{\beta}(s)\right\}.$

Finally, 
we verify the Lindeberg-type condition \citep{van2000}.  
For arbitrary $\varepsilon>0$ and a fixed time point $s\in[h,\tau-h],$ we need to prove
\begin{align*}
 &\lim_{n\to \infty}\frac{1}{n} \sum_{i=1}^n E\Bigg( \left\|\sqrt{h_1h_2}\int_0^{\tau}\int K_{h_1,h_2}(t-s,r-s)\left[{Z}_i(r)-\bar{Z}\{\hat{\beta}(s),t\}\right]\xi_i d N_i^*(r)d N_i(s)\right\|^2 \\
&\times I\left\{ \left\| \sqrt{h_1h_2}\int_0^{\tau}\int  K_{h_1,h_2}(t-s,r-s)\left[{Z}_i(r)-\bar{Z}\{\hat{\beta}(s),t\}\right]\xi_i d N_i^*(r)d N_i(s)  \right\| \geq \sqrt{n}\varepsilon\right\} \Big| D_n\Bigg) \\
&\to 0,
\end{align*}
where $\|x\|=\sqrt{x^Tx},$ for $x=(x_1,x_2,\ldots,x_p)^T.$ Define 
\begin{align*}W&= \frac{1}{n}\sum_{i=1}^n \left\|\sqrt{h_1h_2} \int_0^{\tau}\int K_{h_1,h_2}(t-s,r-s)\left[{Z}_i(r)-\bar{Z}\{\hat{\beta}(s),t\}\right]  d N_i^*(r)d N_i(s)\right\|^2\\
&=\frac{1}{n}\sum_{i=1}^n {W^{\prime}_i}^2.\end{align*}
Then,
\begin{align*}
 & \frac{1}{n} \sum_{i=1}^n E\Bigg( \left\|\sqrt{h_1h_2}\int_0^{\tau}\int K_{h_1,h_2}(t-s,r-s)\left[{Z}_i(r)-\bar{Z}\{\hat{\beta}(s),t\}\right]\xi_i d N_i^*(r)d N_i(s)\right\|^2 \\
&\times I\left\{ \left\| \sqrt{h_1h_2}\int_0^{\tau}\int  K_{h_1,h_2}(t-s,r-s)\left[{Z}_i(r)-\bar{Z}\{\hat{\beta}(s),t\}\right]\xi_i d N_i^*(r)d N_i(s)  \right\| \geq \sqrt{n}\varepsilon\right\} \Big| D_n\Bigg) \\
 & =\frac{1}{n} \sum_{i=1}^n E\Bigg(\xi_i^2 \left\|\sqrt{h_1h_2}\int_0^{\tau}\int K_{h_1,h_2}(t-s,r-s)\left[{Z}_i(r)-\bar{Z}\{\hat{\beta}(s),t\}\right]  d N_i^*(r)d N_i(s)\right\|^2 \\
&\times I\left\{ \left| \xi_i \right| \left\| \sqrt{h_1h_2}\int_0^{\tau}\int  K_{h_1,h_2}(t-s,r-s)\left[{Z}_i(r)-\bar{Z}\{\hat{\beta}(s),t\}\right]  d N_i^*(r)d N_i(s)  \right\| \geq \sqrt{n}\varepsilon\right\} \Big| D_n\Bigg)\\
&\leq \max_{1\leq i \leq n}   E\left[\xi_i^2   I\left\{ \left| \xi_i \right| W_i^{\prime} \geq \sqrt{n}\varepsilon\right\} \big| D_n\right]W \\
&\leq E\left[\xi_1^2   I\left\{ \left| \xi_1 \right| \max_{1\leq i \leq n}W_i^{\prime} \geq \sqrt{n}\varepsilon\right\} \big| D_n\right]W.
\end{align*}
Let $W_n=\xi_1^2 W I\left\{ \left| \xi_1 \right| \max_{1\leq i \leq n}W_i^{\prime} \geq \sqrt{n}\varepsilon\right\}.$ It is clear that $0\leq W_n\leq \xi_1^2 W.$ Under \ref{con:count}, \ref{con:kernel} and \ref{con:covariate},  $\lim_{n\to \infty}E( \xi_1^2 W)=\lim_{n\to \infty}  W= E(  \lim_{n\to \infty}\xi_1^2  W)<\infty.$ 
 Applying Pratt's lemma \citep{pratt1960}, 
\begin{align*}
 \lim_{n\to \infty} E\left[\xi_1^2   I\left\{ \left| \xi_1 \right| \max_{i\leq i \leq n}W_i^{\prime} \geq \sqrt{n}\varepsilon\right\} \Big| D_n\right]W \stackrel{p}{\rightarrow} 0.
\end{align*}
It follows that the perturbation-based estimating equation  $\tilde{U}_n\{\hat{\beta}(s)\}$ is asymptotically normal, and is equivalent to ${U}_n\{\hat{\beta}(s)\}$ conditional on the data $D_n=\{(X_i,\delta_i, Z_i(R_{ik}), R_{ik}, k = 1,\ldots, M_i), i = 1, \ldots, n\}.$
\end{proof}

\section{Additional simulation results}
\label{sec:c}
We present additional simulation results with a non-homogeneous Poisson process for $N^*(\cdot)$ and more settings for $\beta_0(t)$. Specifically, the observation time of longitudinal covariates is assumed to follow a non-homogeneous Poisson process with intensity function $8(0.75+(0.5-t)^2)$. The time-varying coefficient $\beta_0(t)$ is set as $3(0.5-t)^2+0.25$ and $\exp(-2t-0.5).$ Other configurations are the same as in the main paper. The simulation results 
are summarized in Table \ref{tabler1} and \ref{tabler2}.

 \begin{table}[htbp]
\caption{Simulation results with $\beta_0(t)=3(0.5-t)^2+0.25.$ }
\begin{center}\small
\label{tabler1}
\begin{tabular}{llccrccccrccc}
\toprule
  & &  &   & \multicolumn{4}{c}{Censoring\ rate\ is\ $15\%$} &  & \multicolumn{4}{c}{Censoring\ rate\ is\ $35\%$}\\
\cline{5-8}  \cline{10-13}
$s$ & $n$ & $h_1$ & $h_2$ & Bias & SE & SD & CP &  & Bias & SE & SD & CP\\
\midrule
$0.2$   & 400 & $n^{-0.35}$ & $n^{-0.25}$& -0.027   & 0.195 & 0.196 & 94.8 &  &   -0.032    & 0.209 & 0.216 & 94.3\\
        &     & $n^{-0.35}$ & $n^{-0.35}$ & -0.011    & 0.222 & 0.222 & 95.3 &  &  -0.020   & 0.237 & 0.243 & 94.2\\
        &     & auto      &       &  -0.013    & 0.218 & 0.219 & 95.3 &  &  -0.021  & 0.232 & 0.239 & 94.5\\
        & 900 & $n^{-0.35}$ & $n^{-0.25}$ & -0.032     & 0.157 & 0.158 & 95.0 &  & -0.031    & 0.168 & 0.164 & 95.7\\
       &     & $n^{-0.35}$ & $n^{-0.35}$ & -0.019     & 0.184 & 0.187 & 94.7 &  & -0.019    & 0.197 & 0.193 & 95.5\\
          &     & auto    &          & -0.020  & 0.180 & 0.184 & 94.7 &  & -0.021  & 0.192  & 0.190  & 95.4 \\
$0.4$   & 400 & $n^{-0.35}$ & $n^{-0.25}$& -0.022  & 0.238 & 0.246 & 94.6 &  &   -0.019      & 0.276 & 0.288 & 93.5\\
        &     & $n^{-0.35}$ & $n^{-0.35}$ & -0.010    & 0.272 & 0.280 & 94.3 &  &  -0.008  & 0.314 & 0.336 & 92.7\\
 &     & auto      &        &  -0.011   & 0.268 & 0.278 & 94.1 &  &  -0.011  & 0.308 & 0.330 & 93.0\\
        & 900 & $n^{-0.35}$ & $n^{-0.25}$ & -0.015   & 0.192 & 0.193 & 95.0 &  & -0.016  & 0.224 & 0.228 & 94.6\\
        &     & $n^{-0.35}$ & $n^{-0.35}$& -0.008     & 0.226 & 0.228 & 95.2 &  & -0.005    & 0.263 & 0.267 & 93.1\\
           &     & auto      &        &  -0.007  & 0.222 & 0.226 & 94.9 &  &  -0.005  & 0.257 & 0.262 & 93.4\\
$0.6$   & 400 & $n^{-0.35}$ & $n^{-0.25}$&  -0.003  & 0.281 & 0.287 & 93.4 &  &   -0.020   & 0.370 & 0.413 & 91.4\\
         &     & $n^{-0.35}$ & $n^{-0.35}$ &  0.013    & 0.321 & 0.330 & 94.2 &  &  -0.011     & 0.421 & 0.474 & 91.7\\
      &     & auto     &         &  0.013   & 0.316 & 0.328 & 94.0 &  &  -0.014  & 0.413 & 0.467 & 91.5\\
        & 900 & $n^{-0.35}$ & $n^{-0.25}$&  -0.001      & 0.227 & 0.231 & 94.5
        &  & -0.016     & 0.296 & 0.311 & 93.4\\
        &     & $n^{-0.35}$ & $n^{-0.35}$&  0.008     & 0.267 & 0.275 & 93.2 &  &  -0.007   & 0.349 & 0.365 & 94.1\\
           &     & auto    &          &  0.009  & 0.262 & 0.271 & 92.9 &  &  -0.011  & 0.341 & 0.356 & 94.2\\
$0.8$   & 400 & $n^{-0.35}$ & $n^{-0.25}$& -0.040   & 0.334 & 0.362 & 92.6 &  &  -0.026     & 0.488 & 0.551 & 91.1\\

       &     & $n^{-0.35}$ & $n^{-0.35}$ &  -0.022    & 0.376 & 0.407 & 91.8  &  &   -0.000    & 0.546 & 0.636 & 91.2\\
      &     & auto    &          &  -0.023   & 0.371 & 0.403 & 92.0 &  &  0.003  & 0.535 & 0.626 & 91.0\\
        & 900 & $n^{-0.35}$ & $n^{-0.25}$&  -0.033    & 0.267 & 0.282 & 93.5 &  &  -0.015       & 0.394 & 0.418 & 92.8\\
        &     & $n^{-0.35}$ & $n^{-0.35}$&  -0.013   & 0.312 & 0.339 & 92.0 &  &  0.007      & 0.458 & 0.495 & 92.4\\
     &     & auto      &        &  -0.013  & 0.306 & 0.335 & 91.7 &  &   0.002  & 0.447 & 0.487 & 92.5\\
\bottomrule
\end{tabular}
\end{center}
\footnotesize{
Note: ``BD'' represents different bandwidths, ``Bias'' is the difference between $\beta_0(t)$ and $\hat{\beta}(t)$,
``SD'' is the sample standard deviation, ``SE'' is the average of the standard error estimates, ``CP''$/100$ represents the coverage probability of the $95\%$ confidence interval for estimators of $\beta_0(s)$ at fixed $s$, and LVCF represents the last value carried forward method. 
}
\end{table}

\begin{table}[htbp]
\caption{Simulation results with $\beta_0(t)=\exp(-2t-0.5).$ }  
\begin{center}\small
\label{tabler2}
\begin{tabular}{llccrccccrccc}
\toprule
  & &  &   & \multicolumn{4}{c}{Censoring\ rate\ is\ $15\%$} &  & \multicolumn{4}{c}{Censoring\ rate\ is\ $35\%$}\\
\cline{5-8}  \cline{10-13}
$s$ & $n$ & $h_1$ & $h_2$ & Bias & SE & SD & CP &  & Bias & SE & SD & CP\\
\midrule
$0.2$   & 400 & $n^{-0.35}$ & $n^{-0.25}$& -0.029   & 0.115 & 0.111 & 94.9 &  &   -0.034    & 0.122 & 0.127 & 92.3\\
        &     & $n^{-0.35}$ & $n^{-0.35}$ & -0.010    & 0.131 & 0.133 & 94.5 &  &  -0.018   & 0.140 & 0.147 & 92.9\\
        &     & auto      &       &  -0.012    & 0.127 & 0.132 & 94.8 &  &  -0.013  & 0.134 & 0.143 & 92.5\\
        & 900 & $n^{-0.35}$ & $n^{-0.25}$ & -0.032     & 0.093 & 0.094 & 92.6 &  & -0.023    & 0.100 & 0.101 & 93.5\\
       &     & $n^{-0.35}$ & $n^{-0.35}$ & -0.010     & 0.109 & 0.115 & 93.0 &  & -0.020    & 0.116 & 0.122 & 93.1\\
          &     & auto    &          & -0.018   & 0.106 & 0.108 & 93.1 &  & -0.017  & 0.113  & 0.113  & 94.3 \\
$0.4$   & 400 & $n^{-0.35}$ & $n^{-0.25}$& -0.022      & 0.144 & 0.142 & 94.2 &  &   -0.018      & 0.167 & 0.172 & 93.3\\
        &     & $n^{-0.35}$ & $n^{-0.35}$ & -0.011    & 0.164 & 0.161 & 94.6 &  &  -0.012  & 0.193 & 0.210 & 91.0\\
 &     & auto      &        &  -0.008   & 0.160 & 0.163 & 93.5 &  &  -0.010  & 0.187 & 0.192 & 93.0\\
         & 900 & $n^{-0.35}$ & $n^{-0.25}$& -0.018      & 0.116 & 0.119 & 93.5 &  &   -0.012      & 0.137 & 0.139 & 93.3\\
        &     & $n^{-0.35}$ & $n^{-0.35}$ & -0.011    & 0.137 & 0.144 & 93.1 &  &  -0.018  & 0.158 & 0.169 & 92.8\\
           &     & auto      &        &  -0.013   & 0.132 & 0.139 & 93.4 &  &  -0.010  & 0.153 & 0.160 & 93.1\\
$0.6$   & 400 & $n^{-0.35}$ & $n^{-0.25}$&  -0.017  & 0.170 & 0.181 & 92.7 &  &   -0.018   & 0.215 & 0.241 & 91.5\\
         &     & $n^{-0.35}$ & $n^{-0.35}$ &  -0.013    & 0.192 & 0.202 & 92.6 &  &  -0.010     & 0.245 & 0.271 & 91.6\\
      &     & auto     &         &  0.002   & 0.186 & 0.191 & 93.0 &  &  -0.016  & 0.240 & 0.256 & 93.0\\
        & 900 & $n^{-0.35}$ & $n^{-0.25}$&  -0.013      & 0.135 & 0.142 & 92.9
        &  & -0.008     & 0.177 & 0.182 & 94.6\\
        &     & $n^{-0.35}$ & $n^{-0.35}$&  0.004     & 0.161 & 0.166 & 94.6 &  &  -0.004   & 0.206 & 0.226 & 92.1\\
           &     & auto    &          &  0.000  & 0.154 & 0.163 & 92.9 &  &   0.002  & 0.198 & 0.216 & 92.8\\
$0.8$   & 400 & $n^{-0.35}$ & $n^{-0.25}$&  0.002   & 0.189 & 0.199 & 92.2 &  &   0.012     & 0.281 & 0.322 & 90.2\\

       &     & $n^{-0.35}$ & $n^{-0.35}$ &  0.004    & 0.218 & 0.228 & 92.4  &  &  -0.001    & 0.324 & 0.371 & 90.1\\
      &     & auto    &          &  -0.005   & 0.210 & 0.231 & 92.1 &  &  0.014  & 0.308 & 0.355 & 89.5\\
        & 900 & $n^{-0.35}$ & $n^{-0.25}$&  -0.005    & 0.155 & 0.160 & 93.3 &  &  -0.002       & 0.225 & 0.254 & 91.1\\
        &     & $n^{-0.35}$ & $n^{-0.35}$&  -0.012   & 0.180 & 0.190 & 92.6 &  &  -0.006      & 0.262 & 0.291 & 90.5\\
  &     & auto      &        &  0.010  & 0.174 & 0.189 & 91.7 &  &   -0.014  & 0.246 & 0.287 & 90.8\\
\bottomrule
\end{tabular}
\end{center}
\footnotesize{
Note: ``BD'' represents different bandwidths, ``Bias'' is the difference between $\beta_0(t)$ and $\hat{\beta}(t)$,
``SD'' is the sample standard deviation, ``SE'' is the average of the standard error estimates, ``CP''$/100$ represents the coverage probability of the $95\%$ confidence interval for estimators of $\beta_0(s)$ at fixed $s$, and LVCF represents the last value carried forward method. 
}
\end{table}
We observe that the bias is small, the average estimated standard error is close to the empirical standard deviation and the coverage probability is close to the nominal $95\%.$
The performance improves with increased sample size corroborating our asymptotic prediction.

\end{document}